\documentclass[12pt]{article}
\usepackage{amssymb}
\usepackage{mathtools}
\usepackage{amsmath}
\usepackage[colorlinks,linkcolor=red,anchorcolor=black,citecolor=green]{hyperref}
\usepackage{xcolor}
\usepackage{graphicx}
\usepackage{multirow}
\usepackage{array}
\usepackage{cite}
\usepackage{pifont}
\usepackage{mathrsfs}
\usepackage{soul}
\usepackage{hepnames}
\usepackage{subcaption}
\usepackage[T1]{fontenc}
\usepackage{lmodern}

\usepackage{longtable}

\usepackage[normalem]{ulem}

\usepackage{listings}

\usepackage{tabularx}
\usepackage{physics}

\newcommand{\ignore}[1]{}

\def\pslash{p\!\!\!\slash}

\allowdisplaybreaks

\begin{document}

\title{
\begin{flushright}
\hfill\mbox{{\small\tt DESY-22-211}}\\[5mm]
\begin{minipage}{0.2\linewidth}
\normalsize
\end{minipage}
\end{flushright}
\textbf{\large Systematic study of one-loop realizations of $d=7$ long-range $0\nu\beta\beta$ decay operators }}

\date{}

\author{
Ping-Tao Chen$^{1}$\footnote{E-mail: {\tt
chenpt@mail.ustc.edu.cn}}, \
Gui-Jun~Ding$^{1}$\footnote{E-mail: {\tt
dinggj@ustc.edu.cn}},  \
Chang-Yuan~Yao$^{2,3}$\footnote{E-mail: {\tt
yaocy@nankai.edu.cn}}
\\*[20pt]
\centerline{
\begin{minipage}{\linewidth}
\begin{center}
$^1${\it \small
Department of Modern Physics, University of Science and Technology of China,\\
Hefei, Anhui 230026, China}\\[2mm]
$^2${\it \small School of Physics, Nankai University, Tianjin 300071, China}\\[2mm]
$^3${\it \small Deutsches Elektronen-Synchrotron DESY, Notkestr. 85, 22607 Hamburg, Germany}
\end{center}
\end{minipage}}
\\[10mm]}

\maketitle

\begin{abstract}

We study the systematical one-loop decomposition of the dimension-7 long-range $0\nu\beta\beta$ decay operators. We find that there are 3 genuine one-loop topologies and 8 diagrams. The procedure to determine the SM quantum number assignments for both internal and external fields is presented. The Majorana neutrino mass in long-range $0\nu\beta\beta$ models is discussed. We also present a one-loop $0\nu\beta\beta$ decay model which produces Majorana neutrino mass at three-loop level. The phenomenological predictions for light neutrino mass and $0\nu\beta\beta$ decay half-life time including both mass mechanism and long-range contribution are studied.

\end{abstract}
\thispagestyle{empty}
\vfill

\newpage

{\hypersetup{linkcolor=black}
\tableofcontents
}

\section{Introduction}

The nature of neutrinos and the origin of neutrino mass are great puzzles in particle physics. In order to accommodate the tiny neutrino masses, one has to extend the standard model(SM). Without extending the gauge symmetry of SM and introducing additional global symmetry, the light neutrinos prefer to be Majorana particles. If neutrinos are Dirac particles, the corresponding Yukawa couplings would be as small as about $10^{-12}$ and certain gauge/global symmetry such as $U_{B-L}(1)$ is necessary to forbid the Majorana mass terms of right-handed neutrinos. At present, we still don't know whether neutrinos are Majorana or Dirac particles. It is well known that the search for the Standard Model (SM) forbidden neutrinoless double-beta $(0\nu\beta\beta)$ decay is the most practical way to probe the Majorana nature of neutrinos. $0\nu\beta\beta$ decay is a transition from a parent nucleus $(A, Z)$ to a daughter nucleus $(A, Z+2)$ with two electrons accompanied but no neutrinos emitted. Obviously, the lepton number is violated by two units in $0\nu\beta\beta$ decay, hence the searches for $0\nu\beta\beta$ decay are searches for lepton-number violation whose observation would demonstrate the breaking of a global conservation law of the SM. It is usually assumed that the $0\nu\beta\beta$ decay is induced by exchange of light Majorana neutrinos between two charged current vertices, then the decay rate is proportional to the square of the effective Majorana neutrino mass $m_{\beta\beta}=\sum^{3}_{i=1} U^2_{ei}m_{i}$ where $U_{ei}$ denotes the element of the lepton mixing matrix and $m_{i}$ are the light neutrino masses. This is the so-called mass mechanism. The current most stringent constraints on the $0\nu\beta\beta$ decay half-life in $^{136}\text{Xe}$ is provided by the KamLAND-Zen experiment~\cite{KamLAND-Zen:2022tow}:
\begin{equation}
T_{1/2}(^{136}\text{Xe})>2.3\times 10^{26}\text{yr}
\end{equation}
at $90\%$ confidence level. This corresponds to upper limits on the effective Majorana neutrino mass in the range $36~\text{meV}\leq |m_{\beta\beta}|\leq 156~\text{meV}$, where the uncertainties mainly arise from the nuclear matrix elements in different nuclear models. Recent theoretical developments in $0\nu\beta\beta$ decay have revealed a new leading contribution from an transition operator induced by light Majorana neutrinos. The associated low-energy constants could lead to significant uncertainty~\cite{Cirigliano:2018hja,Dekens:2020ttz,Graf:2022lhj}.
Conversely, if $0\nu\beta\beta$ decay is observed, neutrinos must be Majorana particles~\cite{Schechter:1981bd}. However, the $0\nu\beta\beta$ decay could also be induced by other new physics effects beyond that of Majorana neutrino masses. In general, the possible mechanism of $0\nu\beta\beta$ decay can be categorized into two classes: the short-range contributions~\cite{Pas:2000vn} and the long-range contributions~\cite{Pas:1999fc}. The short-range part of the $0\nu\beta\beta$ decay amplitude is mediated by the exchange of heavy particles with masses larger than 100 MeV~\cite{Pas:2000vn}, and it is described by a set of dimension-9 operators at leading order~\cite{Pas:2000vn,Babu:2001ex}. The ultraviolet completions of the short-range operators of $0\nu\beta\beta$ decay has been systematically studied at both tree level~\cite{Bonnet:2012kh} and one-loop level~\cite{Chen:2021rcv}. The long-range contributions are induced by the exchange of a light neutrino between two nucleons. If the interaction vertices of both nucleons are the SM charged current interactions, it is exactly the mass mechanism. The long-range contribution to the $0\nu\beta\beta$ decay can appear in new physics models with lepton number violation~(LNV), such as the R-parity violating supersymmetric models~\cite{Mohapatra:1986su,Hirsch:1995zi,Hirsch:1995ek,Babu:1995vh,Pas:1998nn}, the left-right symmetric models~\cite{Mohapatra:1980yp,Deppisch:2014zta,Borah:2017ldt,Li:2020flq} and the leptoquark models~\cite{Hirsch:1996ye,Choubey:2012ux,Brahmachari:2002xc}. The $0\nu\beta\beta$ decay rate including both short-range and long-range parts  has been studied in the framework of effective field theory~\cite{Cirigliano:2017djv,Cirigliano:2018yza}.

The long-range $0\nu\beta\beta$ decay can be described by dimension-7 lepton number violating operators~\cite{Helo:2016vsi,Deppisch:2017ecm}, the complete tree-level decomposition of these dimension-7 operators which induce momentum enhanced contributions to long-range $0\nu\beta\beta$ decay has been discussed in Ref.~\cite{Helo:2016vsi}. In the present work, we shall give a systematical and complete classification of all models contributing to the $d=7$ operators at one-loop level. The procedures to attach external fields and determine the SM quantum numbers of internal fields are presented. Certain quantum number assignments are excluded by the absence of tree-level diagrams in a genuine one-loop $0\nu\beta\beta$ decay model. The long-range $0\nu\beta\beta$ decay operators violate lepton number by two units, consequently the mediators of any $0\nu\beta\beta$ decay model can generate Majorana neutrino mass. The long-range contribution of one-loop is expected to be subdominant to the mass mechanism without fine tuning of parameter values if the neutrino mass is produced at tree or one-loop level. For models with two-loop or higher-loop level neutrino mass, the long-range contribution can be comparable to or dominant over the mass mechanism.

In this work, a systematic UV completion method is utilized to generate classes of models exhaustively for a given operator based on topologies and diagrams. The resulting UV models provide all necessary information, including the Lorentz nature and the SM quantum numbers of the new particles required for the models, which is sufficient to generate the UV Lagrangian. Since the large number of models are generated, it is not feasible to provide a detailed analysis of each model, such as the scalar potential, the mass spectrum, and other aspects required for a realistic phenomenology analysis of UV completions. However, the comprehensive analysis of UV completions here can at least help to identify models that deserve further study, such as those with fewer new particles or particles with specific properties.

The rest of this paper is organized as follows. We present the effective operators for long-range $0\nu\beta\beta$ decay below and above the electroweak~(EW) scale in section~\ref{sec:0nbb-operators}. The strategy of decomposing the long-range $0\nu\beta\beta$ decay operators at one-loop level is studied in section~\ref{sec:generate}, and we give the procedures of generating the topologies and diagrams and models for long-range $0\nu\beta\beta$ decay. The relation between long-range $0\nu\beta\beta$ decay model and neutrino mass is discussed in section~\ref{sec:relaton to mass}. Dominance of the one-loop long-range contribution over the mass mechanism requires that neutrino mass should be generated at two-loop and higher loop levels. We study one example of a one-loop model in detail in section~\ref{sec:example}, and we discuss the constraints imposed by the half-life times of the isotopes ${}^{76}$Ge and ${}^{136}$Xe. Finally, we summarize and present our conclusions in section~\ref{sec:conclusion}.

\section{\label{sec:0nbb-operators}Effective operators for long-range $0\nu\beta\beta$ decay}

At low energy below the electroweak scale, the most general Lagrangian for the long-range $0\nu\beta\beta$ decay can be written as~\cite{Pas:1999fc,Deppisch:2012nb}:
\begin{equation}
\label{eq:L4-fermion}
\mathcal{L}_{\text{eff}}=\frac{G_{F}}{\sqrt{2}}
\left[j^{\mu}_{V-A}J_{V-A,\mu}+\sum_{\alpha,\beta\neq V-A}\epsilon_{\alpha}^{\beta}j_{\beta}J_{\alpha}
\right]\,,
\end{equation}
where the effective coupling constants $\epsilon_{\alpha}^{\beta}$ are scaled with respect to the SM charged current strength $G_F/\sqrt{2}$. The leptonic (hadronic) currents $j_{\beta}$ ($J_{\alpha}$) are defined as:
\begin{gather}
\nonumber J^{\mu}_{V\pm A}=
\overline{u}\gamma^{\mu}(1\pm\gamma_5)d\,,
\qquad  j_{V\pm A}^{\mu}=\overline{e}\gamma^{\mu}(1\pm\gamma_5)\nu\,,
\\
\nonumber J_{S\pm P}=\overline{u}(1\pm\gamma_5)d\,, \qquad
j_{S\pm P}=\overline{e}(1\pm\gamma_5)\nu\,,\\
J^{\mu\nu}_{T_{R/L}}=\overline{u}\gamma^{\mu\nu}(1\pm\gamma_5)d\,, \qquad
j_{T_{R/L}}^{\mu\nu} =\overline{e}\gamma^{\mu\nu}(1\pm\gamma_5)\nu\,, \label{eq:long-range-0nbb-EF-ope}
\end{gather}
with $\gamma^{\mu\nu}=\frac{i}{2}[\gamma^{\mu}, \gamma^{\nu}]$ and $\nu\equiv\nu_{L}+\nu_{L}^{c}$, where $\nu_{L}^{c}=\mathcal{C}\overline{\nu_L}^T$ is the charge conjugation field of left-handed neutrino and $\mathcal{C}$ is the charge conjugation matrix. We can see that  all currents involving operators proportional to $(1+\gamma_{5})$ will pick the component $\nu_{L}^{C}$ and consequently they would violate lepton number by two units. In Eq.~\eqref{eq:L4-fermion}, one should sum over all possible contractions of leptonic and hadronic currents allowed by Lorentz-invariance, in other words, all possible combinations of Lorentz indices $\alpha, \beta$ should be considered. Notice the identity $j_{T_{R}}^{\mu\nu}J_{T_{L}\mu\nu}=j_{T_{L}}^{\mu\nu}J_{T_{R}\mu\nu}=0$, consequently there are only ten independent operators in Eq.~\eqref{eq:L4-fermion} for the long-range $0\nu\beta\beta$ decays.

\begin{figure}[t!]
\centering
\includegraphics[width=0.85\textwidth]{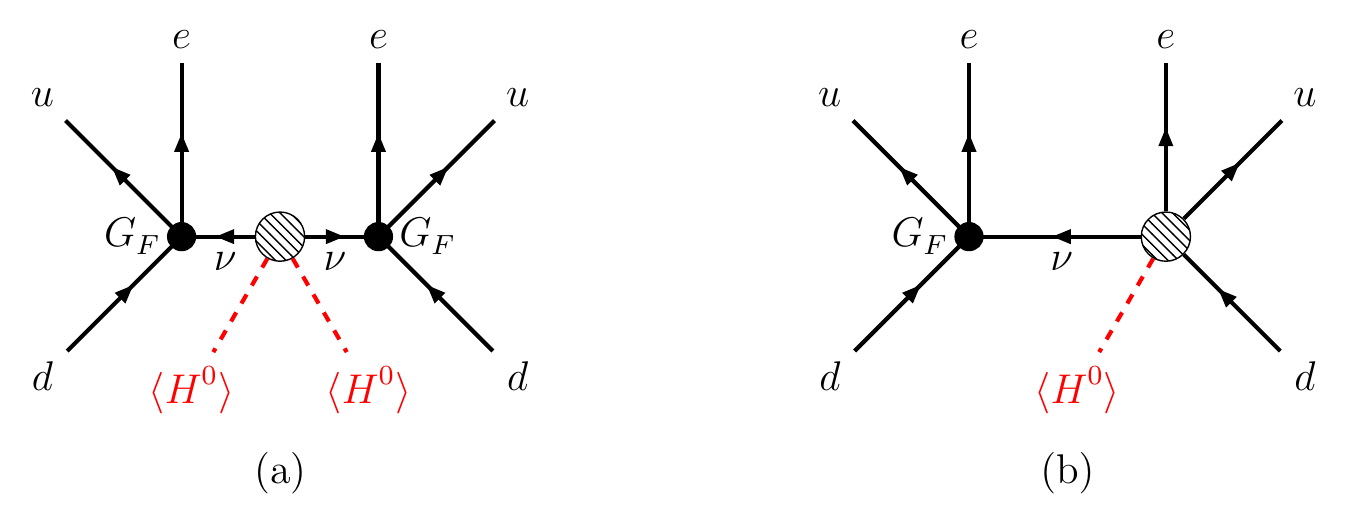}
\caption{\label{fig:LRGenaral}The mass mechanism (left panel) and long-range contributions (right panel) to the $0\nu\beta\beta$ decay rate, where the black dot denotes the SM effective four-fermion interaction, while the slashed circle stands for the effective vertex of the neutrino masses (left panel) and the long-range $0\nu\beta\beta$ operator (right panel) arising from new physics. Here we do not show the diagram with two new physics vertices, since the corresponding contribution is suppressed. }
\end{figure}

The long-range part of $0\nu\beta\beta$ decay is induced by the exchange of a light neutrino between two point-like vertices. If both interaction vertices are the SM charged current interactions, it yields the mass mechanism.
If both interaction vertices are new physics contributions parameterized by Eq.~\eqref{eq:L4-fermion}, the corresponding amplitudes would be quadratic in $\epsilon_{\alpha}^{\beta}$ and they are too small to be negligible\footnote{The contribution of the new interaction relies on the specific $0\nu\beta\beta$ decay mechanism. The new contribution possibly dominate over the standard mechanism, even with two new interaction vertices but meditated by particles other than light neutrinos~\cite{Dekens:2020ttz}. However, in this work, we choose to focus on the scenario of introducing only one new interaction vertex, as it is the leading order new effect.}.
In the present work, we shall be concerned with the case that only one vertex arises from the new physics beyond SM and the other one is the SM charged current interaction, as shown in figure~\ref{fig:LRGenaral}. Then the $0\nu\beta\beta$ decay amplitude is proportional to the time-ordered product of the Lagrangian of the two interaction vertices
\begin{equation}
\int\;d^4x\int\;d^4y\;\frac{G^2_F}{2}\epsilon_{\alpha}^{\beta}\;\mathcal{T}\left[j_{\beta}(x)J_{\alpha}(x)j^{\mu}_{V-A}(y)J_{V-A,\mu}(y)\right]\,.
\end{equation}

If the non-SM lepton current $j_{\beta}$ is left-handed with $\beta=(S-P), T_L$, lepton number violation arises from the Majorana mass terms of light neutrinos. Then the light neutrino mass in the numerator of the neutrino propagator would be picked out by the chiral projection operator $1\pm\gamma_5$, consequently the resulting amplitude would be proportional to $\epsilon_{\alpha}^{\beta}\langle m_{\nu}\rangle$ which is subdominant to the mass mechanism contribution, where $\langle m_{\nu}\rangle$ is the effective Majorana neutrino mass. On the other hand, if the lepton current $j_{\beta}$ is right-handed with $\beta=(S+P), (V+A), T_R$, the lepton number is violated at the new interaction vertex, and the term $\pslash$ will be projected out from the numerator of the neutrino propagator, where the neutrino momentum is of order $\mathcal{O}(100)$ MeV. As a result, the long-range amplitude is proportional to $\epsilon_{\alpha}^{\beta}p$ and it could be comparable to the standard mass mechanism. In the following sections, we shall study the ultraviolet completion of the lepton number violating long-range $0\nu\beta\beta$ decay operators with right-handed leptonic current: $j_{S+P}J_{S+P}$, $j_{S+P}J_{S-P}$, $j^{\mu}_{V+A}J_{V+A,\mu}$, $j^{\mu}_{V+A}J_{V-A,\mu}$ and $j_{T_{R}}^{\mu\nu}J_{T_{R}, \mu\nu}$. These five operators satisfy the electromagnetic $U(1)$ gauge symmetry, but they are not invariant under the action of the SM gauge group $SU(3)_C\times SU(2)_L\times U(1)_Y$. They arise from the following dimension-7 SM gauge invariant operators~\cite{Babu:2001ex,Lehman:2014jma,Helo:2016vsi},
\begin{align}
\nonumber\mathcal{O}_{1}&\equiv\epsilon^{ik}\epsilon^{jl}
(\overline{\ell^{c}_{i}}\ell_{j})(\overline{d_{R}}Q_{k})H_{l}\,,\\
\nonumber\mathcal{O}_{2}&\equiv\epsilon^{ik}\epsilon^{jl}
(\overline{\ell^{c}_{i}}\gamma^{\mu\nu}\ell_{j})(\overline{d_{R}}\gamma_{\mu\nu}Q_{k})H_{l}\,,\\
\nonumber\mathcal{O}_{3}&\equiv\epsilon^{jk}(\overline{\ell^{c}_{i}}\ell_{j})(\overline{Q}^{i}u_{R})H_{k}\,, \\
\label{eq:onbb-SMEFT}\mathcal{O}_{4}&\equiv(\overline{\ell^{c}_{i}}\gamma^{\mu}e_{R})(\overline{d_{R}}\gamma_{\mu}u_R)\epsilon^{ij}H_{j}\,,
\end{align}
where $i, j, k, l=1, 2$ are the indices of the $SU(2)_L$ gauge group, $\ell=\left(\nu_{eL}, e_L\right)^{T}$ and $Q=\left(u_L, d_L\right)^{T}$ denote the first generation of lepton and quark doublet respectively, $u_R, d_R$ and $e_R$ are the first generation of right-handed quark and lepton singlets, $H$ is the SM Higgs doublet. After the electroweak symmetry breaking by the vacuum expectation value~(VEV) of the Higgs field, the operators $\mathcal{O}_{1}$, $\mathcal{O}_{2}$, $\mathcal{O}_{3}$ and $\mathcal{O}_{4}$ give rise to the low energy long-range $0\nu\beta\beta$ decay operators $j^{\dagger}_{S+P}J^{\dagger}_{S+P}$, $j_{T_{R}}^{\mu\nu\dagger}J^{\dagger}_{T_{R}, \mu\nu}$, $j^{\dagger}_{S+P}J^{\dagger}_{S-P}$ and $j^{\mu\dagger}_{V+A}J^{\dagger}_{V+A,\mu}$ respectively. Notice that the remaining long-range operator $j^{\mu}_{V+A}J_{V-A,\mu}$ is generated by the following dimension-9 SM effective operator~\cite{Liao:2020jmn}
\begin{equation}
\mathcal{O}_{5}=\epsilon^{im}\epsilon^{kn}(\overline{\ell^{c}_i}\gamma^{\mu}e_R)(\overline{Q}^j\gamma_{\mu}Q_{k})H_{m}H_jH_n\,,
\end{equation}
which leads to $0\nu\beta\beta$ decay at higher dimension, we will not discuss this scenario in our current work. Including the lepton flavor indices in Eq.~\eqref{eq:onbb-SMEFT}, one can obtain all independent dimension-7 lepton number violating operators without derivative~\cite{Helo:2016vsi}:
\begin{align}
\nonumber\mathcal{O}_{1}(\alpha,\beta)&\equiv\epsilon^{ik}\epsilon^{jl}
(\overline{\ell^{c}_{\alpha i}}\ell_{\beta j})(\overline{d_{R}}Q_{k})H_{l}\,,\\
\nonumber\mathcal{O}_{2}(\alpha,\beta)&\equiv
\epsilon^{ik}\epsilon^{jl}(\overline{\ell^{c}_{\alpha i}}\gamma^{\mu\nu}\ell_{\beta j})(\overline{d_{R}}\gamma_{\mu\nu}Q_{k})H_{l}\,,\\
\nonumber\mathcal{O}_{3}(\alpha,\beta)&\equiv\epsilon^{jk}
(\overline{\ell^{c}_{\alpha i}}\ell_{\beta j})(\overline{Q}^{i}u_{R})H_{k}\,,\\
\mathcal{O}_{4}(\alpha,\beta)&\equiv(\overline{\ell^{c}_{\alpha i}}\gamma^{\mu}e_{R\beta})(\overline{d_{R}}\gamma_{\mu}u_R)\epsilon^{ij}H_{j}\,,
\end{align}
besides $\epsilon^{ik}\epsilon^{jl}(\overline{\ell^{c}_{\alpha i}}\ell_{\beta j})H_{k}H_{l}(H^{\dagger}H)$ which is the famous Weinberg operator with the addition $H^{\dagger}H$. Here $\alpha$ and $\beta$ are lepton flavor indices. In the following we will study the one-loop decomposition of the $0\nu\beta\beta$ decay operators in Eq.~\eqref{eq:onbb-SMEFT} and the relation with light neutrino mass.

\section{\label{sec:generate}Systematical one-loop decomposition}

In the following, we will use the diagrammatic method~\cite{Antusch:2008tz,Gavela:2008ra} to find out all possible one-loop decomposition of the dimension seven long-range $0\nu\beta\beta$ decay operators in Eq.~\eqref{eq:onbb-SMEFT}. This method has been used to decompose the neutrino mass operators for both Majorana neutrinos~\cite{Bonnet:2012kz,Sierra:2014rxa,Cepedello:2018rfh} and Dirac neutrinos~\cite{Yao:2017vtm,Yao:2018ekp,CentellesChulia:2019xky,Jana:2019mgj}. Firstly, we identify the one-loop topologies with five external legs by using only 3-point vertices and 4-point vertices. The topologies of tadpole and self-energies are eliminated. In the next step, we promote topologies to diagrams by specifying the Lorentz nature (spinor or scalar) of each internal and external lines. Renormalizability and Lorenz invariance require the diagrams contain only the interaction vertices of the type fermion-fermion-scalar, scalar-scalar-scalar or scalar-scalar-scalar-scalar. A topology can lead to a few number of Feynman diagrams, because there are usually several possible assignments of quark fields, lepton fields and Higgs field to the five external legs. Furthermore, each interaction vertex should be invariant under the SM gauge group $SU(3)_C\times SU(2)_L\times U(1)_Y$ such that one can constrain the quantum numbers of the internal fermion and scalar fields. If the gauge quantum numbers of all internal fields are specified for a diagram, the corresponding UV completion will be called a model, and the gauge invariant interactions involving the beyond SM fields can be read out straightforwardly. Notice that the SM gauge quantum numbers of the new fields can be unambiguously fixed in the tree-level realizations, while there are infinite possible quantum number assignments to the fields running in the loop. In the following, we will consider the scenarios that new fields are singlets, doublets or triplets of $SU(2)_L$, and the results for the higher dimensional representations can be derived in a similar way. Regarding $SU(3)_C$ assignment for the fields, the low-dimensional representations up to octet are considered for illustration.

\subsection{\label{sec:Gtopologies}One-loop topologies for long-range $0\nu\beta\beta$ decay operators}

As shown in Eq.~\eqref{eq:onbb-SMEFT}, we see the dimension-7 long-range $0\nu\beta\beta$ decay operators involve two quark fields, two lepton fields and a Higgs doublet. Using our own codes unpublicized yet, we plot the connected one-loop topologies with five external legs, and we find there are 37 one-loop topologies. However, most of the topologies are of no interest to us, we can exclude a lot of them at the topology level. The first step is to exclude all the topologies with tadpoles and self-energy, because the models generated from these topologies always have divergent parts in their loop integrals, and there should be a lower order counter term required by renormalizability.

\begin{figure}[htbp]
\centering
\includegraphics[width=\textwidth]{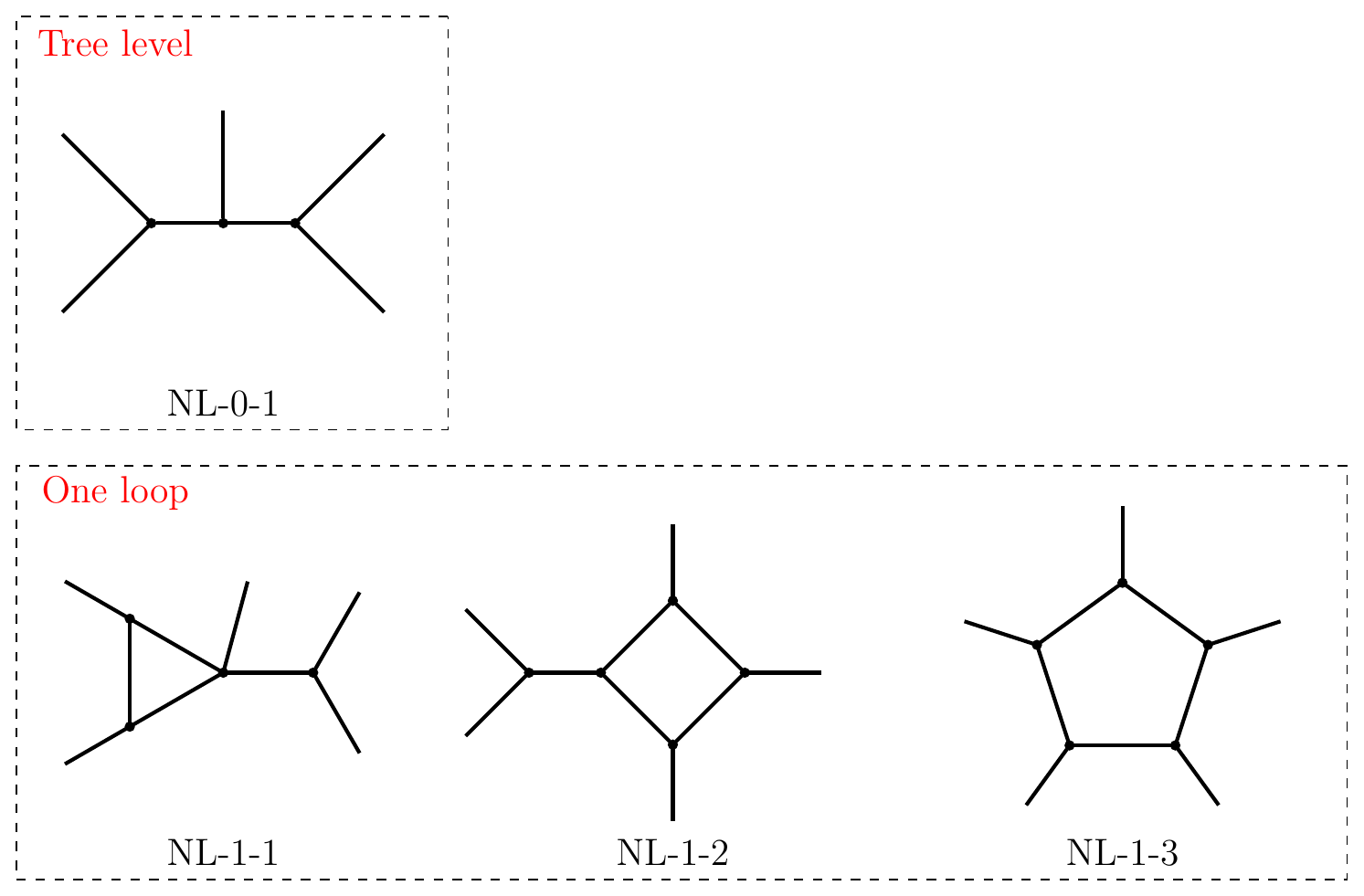}
\caption{The tree-level and one-loop topologies that can lead to genuine models of long-range $0\nu\beta\beta$ decays. }
\label{fig:N0-1}
\end{figure}

\begin{figure}[htbp]
\centering
\includegraphics[width=0.8\textwidth]{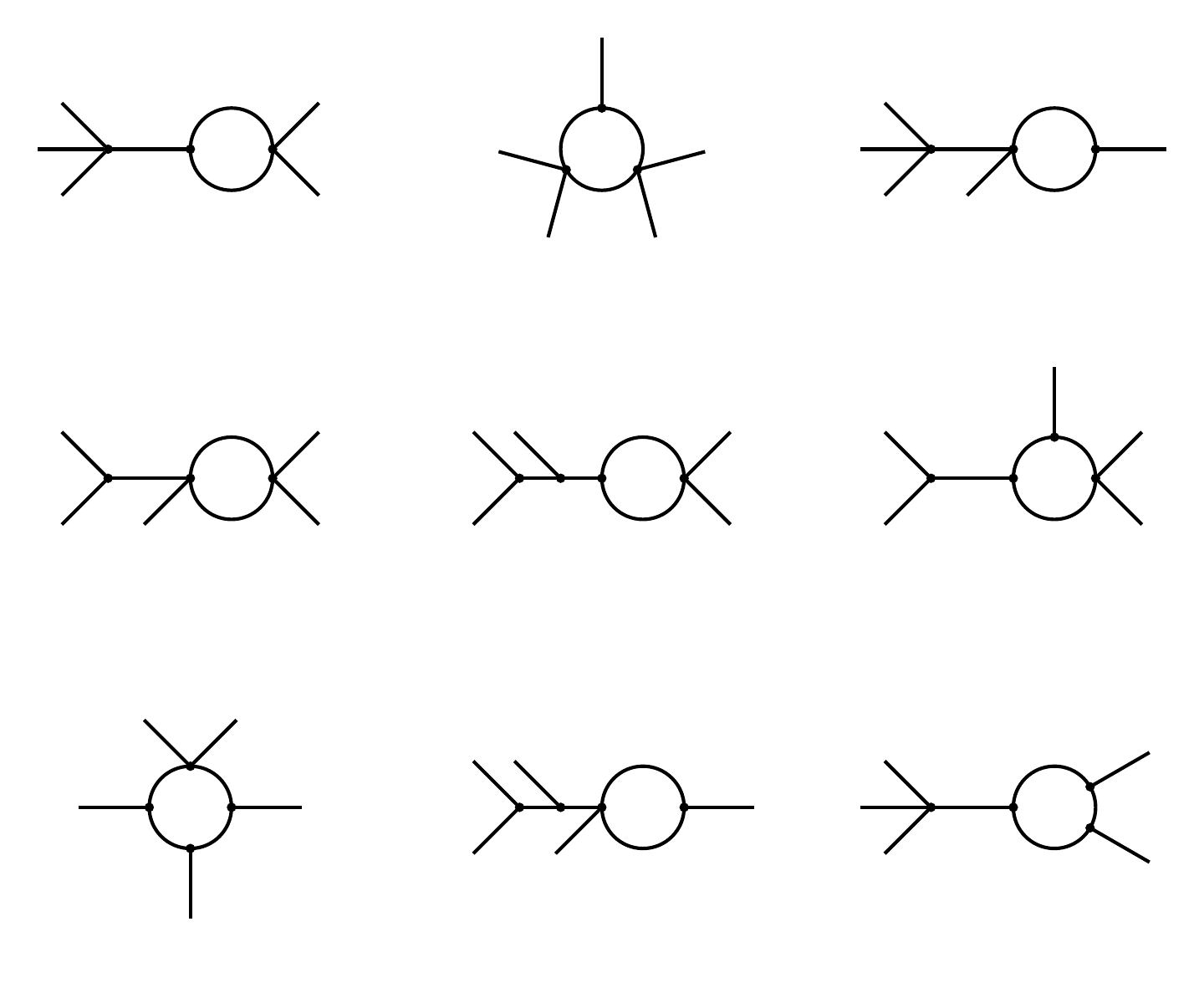}
\caption{The one-loop topologies that always lead to non-renormalizable diagrams.}
\label{fig:NNormalize}
\end{figure}

\begin{figure}[htbp]
\centering
\includegraphics[width=0.5\textwidth]{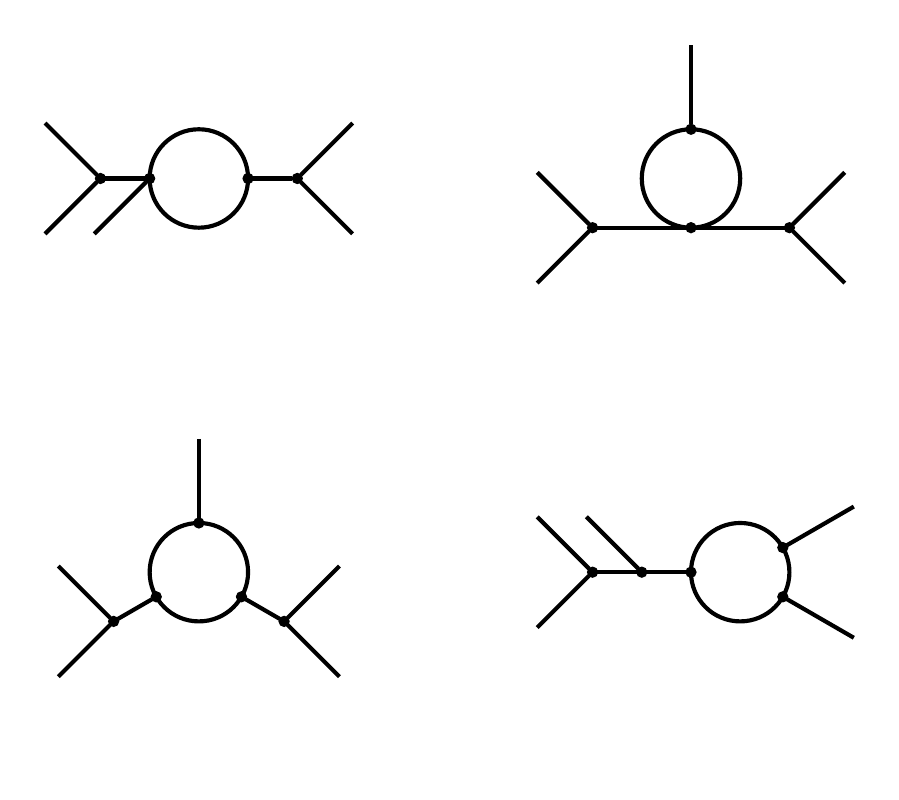}
\caption{The one-loop topologies leading to non-genuine finite or divergent diagrams, the internal loop with three legs can be compressed into a three-point vertex.}
\label{fig:NGenuineT}
\end{figure}

Then there are 16 topologies left. Since we are working on the operators with four fermions and one scalar, some topologies need non-renormalizable interactions to accommodate these external lines. As a consequence, these topologies should be discarded and they are shown in figure~\ref{fig:NNormalize} for completeness. At this point we are left with 7 different topologies. We intend to identify the topologies and diagrams as well as the models for which the leading order contribution to $0\nu\beta\beta$ decay arises at one-loop level, and the tree-level contribution is absent without the need to introduce extra symmetry. These topologies, diagrams and models will be considered genuine. If a diagram has a sub-diagram with a loop and three external legs, then the three-point vertex without the loop is also compatible with the symmetry. In other words, any internal loop (or loop) with three legs can be compressed into a three-point vertex. Thus the corresponding one-loop diagram must be accompanied by the more important tree-level diagram, and consequently it is non-genuine and should be discarded. The topologies with compressible one-loop sub-diagram are displayed in figure~\ref{fig:NGenuineT} for completeness, they can be regarded as extensions of the tree-level topology, where one of the vertices is generated at one-loop. Discarding the compressible topologies in figure~\ref{fig:NGenuineT},  there remain only 3 genuine topologies shown in figure~\ref{fig:N0-1}. We also display the unique tree-level topology in figure \ref{fig:N0-1}, and the systematic decomposition of tree-level $0\nu\beta\beta$ decay model dominated by long-range contribution has been studied in Ref.~\cite{Helo:2016vsi}. In the present work, we also provide the tree-level decomposition, since these results are necessary when determining the genuineness of a one-loop model.

\subsection{\label{sec:Gdiagram}Constructing diagrams}

\begin{figure}[htbp]
\centering
\includegraphics[width=\textwidth]{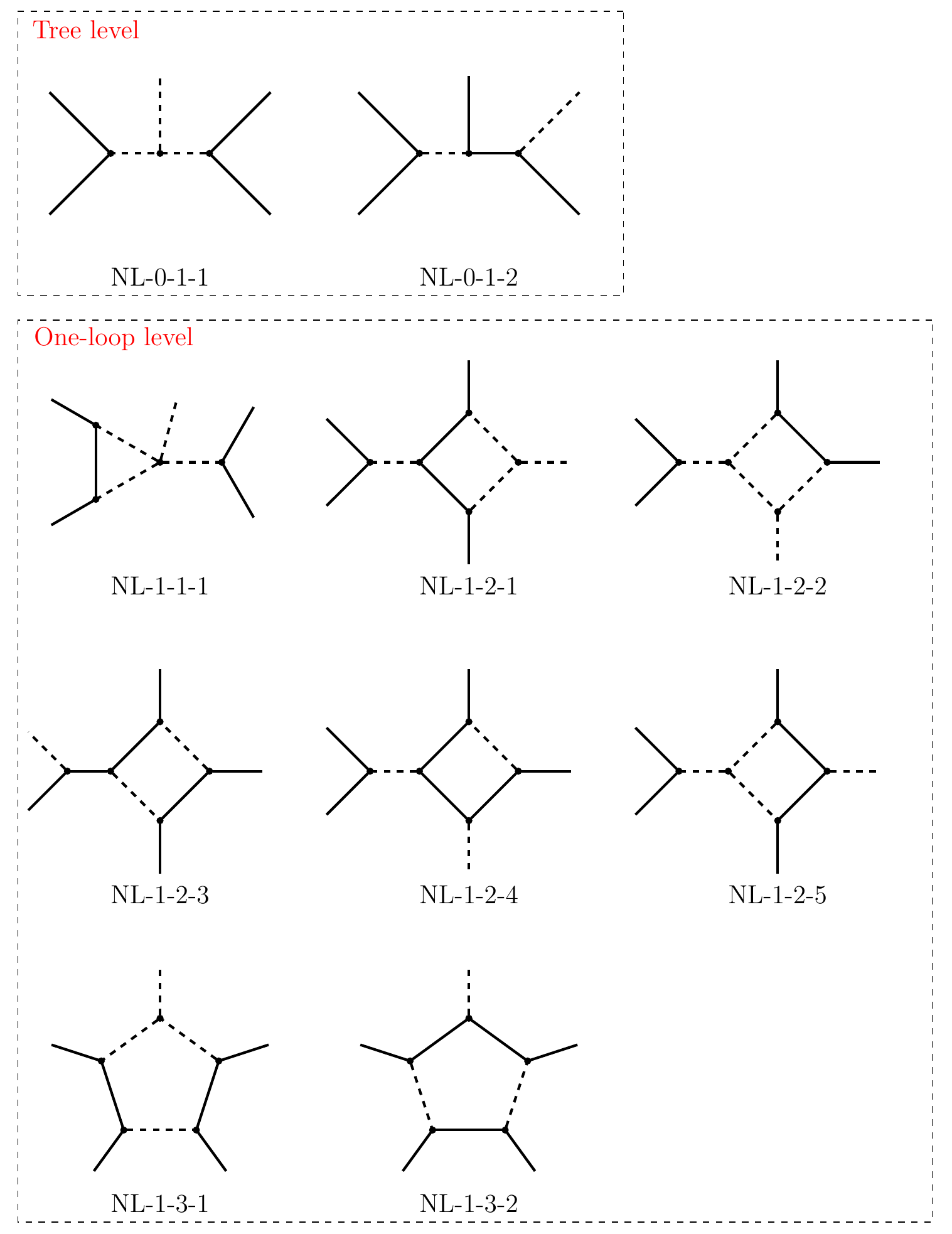}
\caption{List of genuine diagrams for long-range $0\nu\beta\beta$ decay up to one-loop level.}
\label{fig:DiaN0-1}
\end{figure}

\begin{figure}[htbp]
\centering
\includegraphics[width=0.9\textwidth]{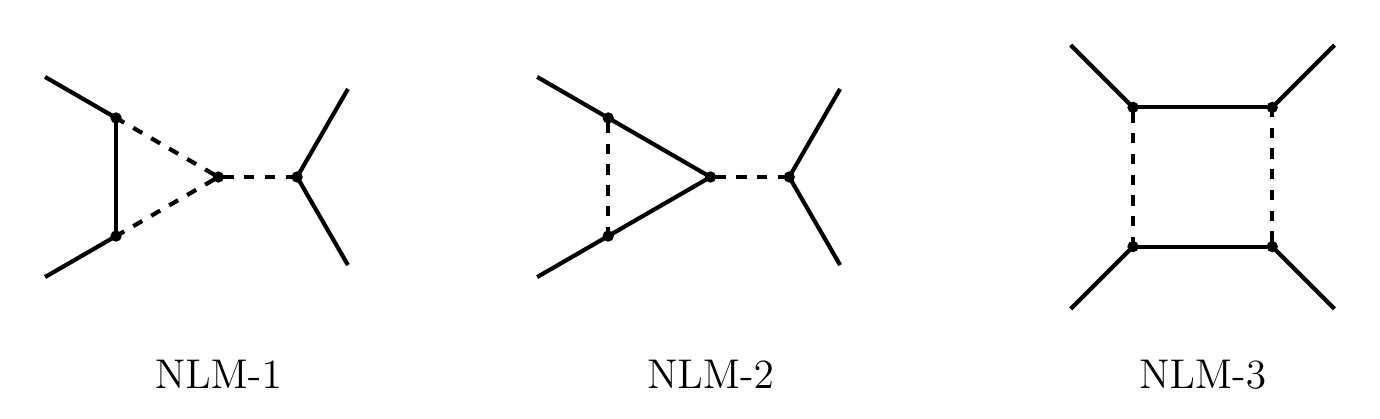}
\caption{The genuine one-loop diagrams for long-range $0\nu\beta\beta$ decay in the mass basis. Notice that the external leg of Higgs is removed after electroweak symmetry breaking.}
\label{fig:MassN1}
\end{figure}

We proceed to specify the Lorentz nature (fermion or scalar) of both external and internal lines of each topology. The SM invariant $0\nu\beta\beta$ decay operators in Eq.~\eqref{eq:onbb-SMEFT} involve two quark fields, two lepton fields and a Higgs field. There are several options for the assignments of the four fermions and one scalar to the five external legs for each topology. After considering all possible external leg assignments,  we insert the fermion or scalar into internal lines one by one and Lorentz invariance implies that each vertex must contain an even number of fermions. The UV completion models are required to be renormalizable so that the dimension of each interaction vertex should be less than or equal to 4. As a consequence, only the renormalizable scalar-scalar-scalar, fermion-fermion-scalar and  scalar-scalar-scalar-scalar interactions can be used. As shown in figure~\ref{fig:DiaN0-1}, we find there are 8 independent one-loop diagrams arising from the genuine topologies of figure~\ref{fig:N0-1}. After the electroweak symmetry breaking, the couplings to external Higgs field lead to chirality flip of fermion field or scalar mixing. The external Higgs field with vacuum expectation value insertion can be removed. Hence the 8 diagrams in figure~\ref{fig:DiaN0-1} get reduced to only 3 diagrams in mass basis, as shown in figure \ref{fig:MassN1}, these diagrams will be useful when we calculate the $0\nu\beta\beta$ decay rate.

\subsection{\label{sec:Gmodel}The approach of generating models}

The next step is to generate models based on the 8 genuine diagrams listed in figure~\ref{fig:DiaN0-1}. We need to specify how each internal line transforms under the SM gauge group $SU(3)_{C}\times SU(2)_{L}\times U(1)_{Y}$. We firstly attach the fields of the effective operators to external lines of the diagram. Subsequently imposing gauge invariance of the interaction vertices, we can determine the possible quantum numbers of the messenger fields. In the following, we give the details of generating models.

\subsubsection{Attaching external fields}

The effective operators $\mathcal{O}_{1,2,3,4}$ of long-range $0\nu\beta\beta$ decay can be classified into three categories according to the fields involved, as summarized in table~\ref{tab:NL1toNL3}, where the conjugate operators are considered to accommodate our convention. We see that $\mathcal{O}_1$ and $\mathcal{O}_2$ are composed of the same fields, therefore they share the same routine of UV completion. Generally both of them are generated in a concrete UV model after integrating out the heavy fields. Each operator class in table~\ref{tab:NL1toNL3} involves different external legs, as a result, the UV completions of these three classes should be performed separately based on the diagrams given in figure~\ref{fig:DiaN0-1}. In the following, we take the diagram NL-1-1-1 with operator $\mathcal{O}_4$ as an example to illustrate the external field assignment. One can decompose the $0\nu\beta\beta$ decay operators for other diagrams in the same fashion. The full results are collected in the attached Mathematica file~\cite{Chen:2021sup2}.

\begin{table}[hptb]
\renewcommand{\tabcolsep}{0.5mm}
\renewcommand{\arraystretch}{1.3}
\centering
\begin{tabular}{|c|c|c|}\hline\hline
\texttt{Name}   ~&~ $0\nu\beta\beta$  decay   operators
~&~     \texttt{External fields}\\ \hline
NL1     &       $\mathcal{O}^{\dagger}_{1}, \mathcal{O}^{\dagger}_{2}$  & $\overline{\ell},\overline{\ell},\overline{Q},d_{R},H^{\dagger}$\\ \hline
NL2     &   $\mathcal{O}^{\dagger}_{3}$  & $\overline{\ell},\overline{\ell},Q,\bar{u}_{R},H^{\dagger}$\\ \hline
NL3     &   $\mathcal{O}^{\dagger}_{4}$  & $\overline{\ell},\bar{e}_{R},\bar{u}_{R},d_{R},H^{\dagger}$\\ \hline\hline
\end{tabular}
\caption{The fields involved in the $0\nu\beta\beta$ decay operators, the hermitian conjugate operators are used to accommodate our convention. }
\label{tab:NL1toNL3}
\end{table}

\begin{figure}[htbp]
\centering
\includegraphics[width=0.9\textwidth]{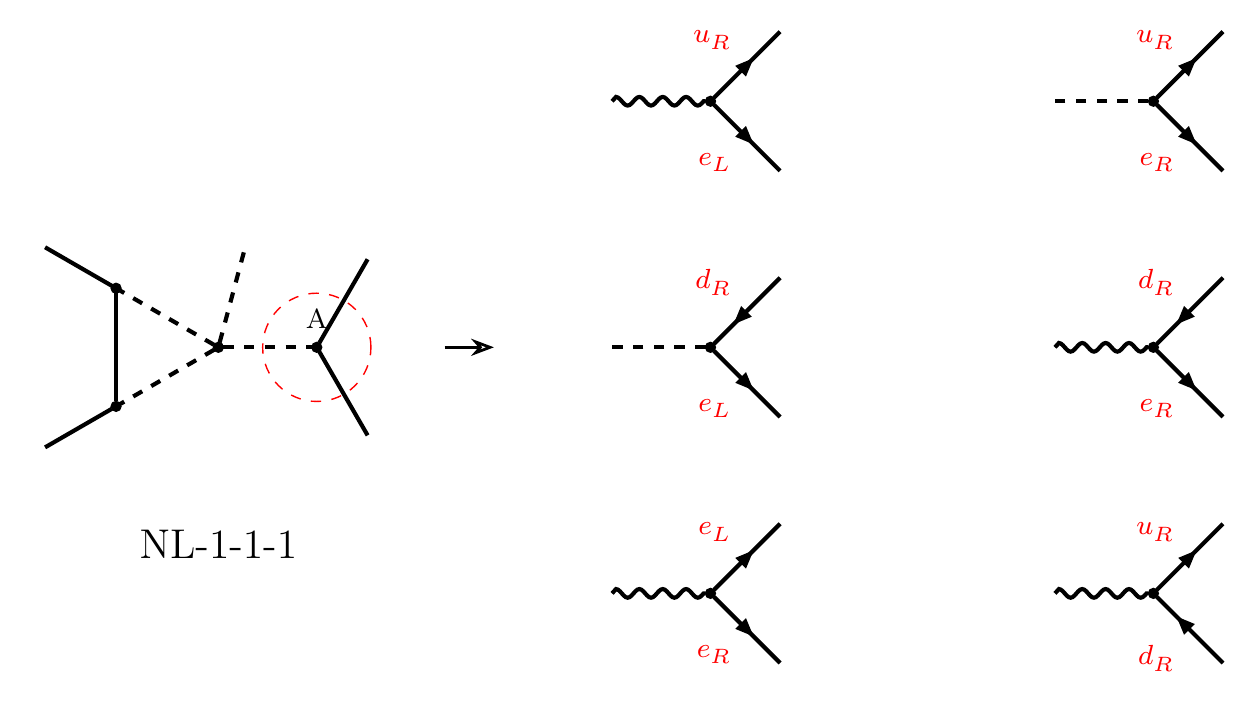}
\caption{Attach the fields of the $0\nu\beta\beta$ decay operator $\mathcal{O}^{\dagger}_4$ to the external lines of the diagram NL-1-1-1 at the vertex $A$. On the right side, the dashed lines stand for scalar fields while the wavy lines denote vector fields. Note that the chirality of the external fermions fixes the mediator to be either vector or scalar boson. }
\label{fig:external}
\end{figure}

\begin{figure}[htbp]
\centering
\includegraphics[width=0.9\textwidth]{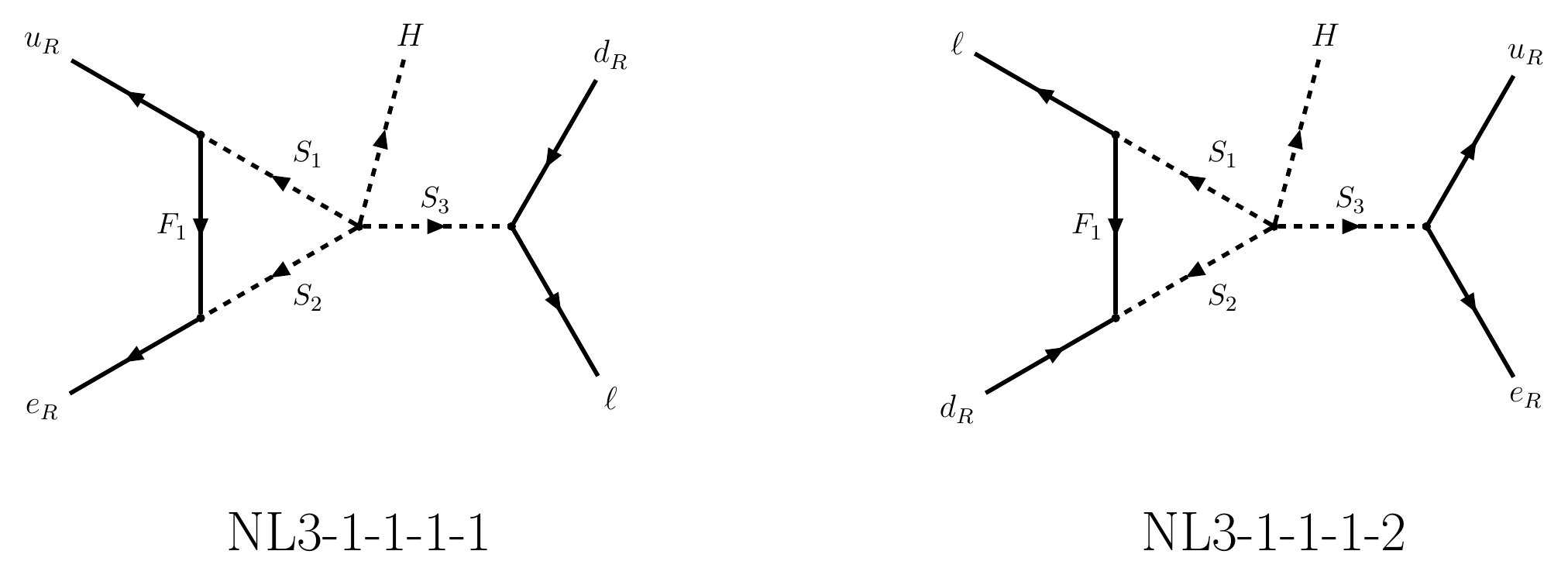}
\caption{Attach external fields of NL3 to the diagram NL-1-1-1. After attaching external fields, we change the notation ``NL-'' to corresponding operator notation ``NL3-''.}
        \label{fig:external-NL3}
\end{figure}

The operator $\mathcal{O}^{\dagger}_4$ is constituted by the fields $\overline{\ell}$, $\bar{e}_{R}$, $\bar{u}_{R}$, $d_{R}$ and $H^{\dagger}$, which can be freely assigned to the external legs. However, the lepton and quark fields in the SM are chiral fields in weak basis, consequently, for certain attachment of external fields, Lorentz invariance requires vector mediators otherwise the vertex would be vanishing. As far as we know,  vector bosons should be the gauge bosons of certain gauge symmetry and their masses are generated through the spontaneous breaking of the extended gauge symmetry. Thus the new gauge bosons require extending both the SM gauge group and scalar field content. In the present work, we would like to preserve the SM gauge group $SU(3)_{C}\times SU(2)_{L}\times U(1)_{Y}$ that has been tested by lots of experiments from low energy to TeV scale. Hence the cases of vector mediators will not be considered\footnote{The scalar mediators can also be the SM gauge bosons if they transforms as $(1, 3, 0)$ or $(1, 1, 0)$ under the SM gauge group $SU(3)_{C}\times SU(2)_{L}\times U(1)_{Y}$ and the relevant fermion-fermion-vector interaction is allowed by the chirality of external fermions. We would like to mention that the results for vectors can be straightforwardly derived from the corresponding ones for
scalars. However, the interaction vertices and the propagator of a massive vector boson are different from those of a scalar, the vector mediator and scalar mediator lead to different contributions.}. Let us consider the attachment of external legs at the vertex $A$ of diagram NL-1-1-1, as shown in figure \ref{fig:external}. Lorentz invariance implies that 4 out of the 6 possible assignments need new vector mediators, and consequently they are discarded. Moreover, one can freely attach the fields to all the external lines, and some attachments are superfluous. In order to identify the redundant ones, we should consider permutations of vertices and compare the different couplings. If two different attachments are related with each other through permutation of vertices, they will be essentially the same one. After attaching all the fields of the operator $\mathcal{O}_4^{\dagger}$ to the external lines of the diagram NL-1-1-1, we find there are only two possible assignments shown in figure~\ref{fig:external-NL3}. Following the above procedure, we have found out all the possible independent external fields attachments for the genuine diagrams in figure~\ref{fig:DiaN0-1} , and the same procedure can be applied to all other $0\nu\beta\beta$ decay operators. Our results are summarized in table~\ref{tab:NL-1ex}, table~\ref{tab:NL-2ex} and table~\ref{tab:NL-3ex}. Once the external lines are specified, the SM quantum numbers of the internal fields can be determined. See the following sections for details.

\begin{table}[htbp]
\renewcommand{\tabcolsep}{0.3mm}
\renewcommand{\arraystretch}{1.2}
\centering
\begin{tabular}{|c|c|c|c|c|c|c||c|c|c|c|c|c|c|}
\hline\hline
\multicolumn{14}{|c|}{\includegraphics[width=0.3\textwidth]{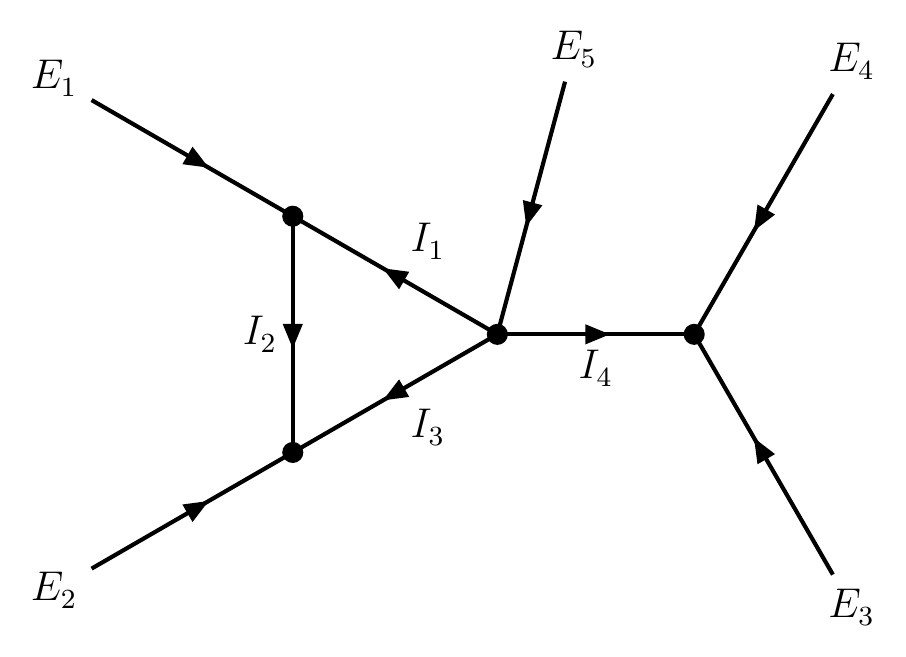}}\\ \hline
Operator & Diagram & $E_{1}$ & $E_{2}$ & $E_{3}$ & $E_{4}$ & $E_{5}$ &
Operator & Diagram & $E_{1}$ & $E_{2}$ & $E_{3}$ & $E_{4}$ & $E_{5}$ \\\hline
\multirow{4}{*}{NL1} & \multirow{4}{*}{NL-1-1-1} & $\overline{\ell}$ & $d_{R}$ & $\overline{\ell}$ & $\overline{Q}$ & $H^{\dagger}$  &
\multirow{2}{*}{NL2} & \multirow{2}{*}{NL-1-1-1} & $\bar{u}_{R}$ & $Q$ & $\overline{\ell}$ & $\overline{\ell}$ & $H^{\dagger}$ \\ \cline{3-7}\cline{10-14}
& & $\overline{Q}$ & $\overline{\ell}$ & $\overline{\ell}$ & $d_{R}$ & $H^{\dagger}$ &  & &$\overline{\ell}$  & $\overline{\ell}$ & $\bar{u}_{R}$ & $Q$ & $H^{\dagger}$ \\\cline{3-14}
& & $\overline{Q}$ & $d_{R}$ & $\overline{\ell}$ & $\overline{\ell}$ & $H^{\dagger}$ & \multirow{2}{*}{NL3} & \multirow{2}{*}{NL-1-1-1} & $\bar{u}_{R}$ & $\bar{e}_{R}$ & $\overline{\ell}$  & $d_{R}$ & $H^{\dagger}$ \\ \cline{3-7}\cline{10-14}
& & $\overline{\ell}$ & $\overline{\ell}$ & $\overline{Q}$ & $d_{R}$ & $H^{\dagger}$ &  & & $\overline{\ell}$ & $d_{R}$ & $\bar{e}_{R}$  & $\bar{u}_{R}$ & $H^{\dagger}$ \\\hline\hline
\end{tabular}
\caption{The possible external field attachments for the topology NL-1-1, where the external and internal fields are labelled as $E_i$ and $I_i$ respectively.}
\label{tab:NL-1ex}
\end{table}

%%%%%%%%%%%%%%%%
%\clearpage
\renewcommand*{\tabcolsep}{0.3mm}
\renewcommand*{\arraystretch}{1.2}
\begin{longtable}{|c|c|c|c|c|c|c||c|c|c|c|c|c|c|}
\hline\hline
\endfirsthead

\hline
Operator & Diagram & $E_{1}$ & $E_{2}$ & $E_{3}$ & $E_{4}$ & $E_{5}$ &
Operator & Diagram & $E_{1}$ & $E_{2}$ & $E_{3}$ & $E_{4}$ & $E_{5}$ \\\hline
\endhead

\hline \multicolumn{14}{|r|}{Continued on next page} \\ \hline
\endfoot

\endlastfoot
\multicolumn{14}{|c|}{\includegraphics[width=0.4\textwidth]{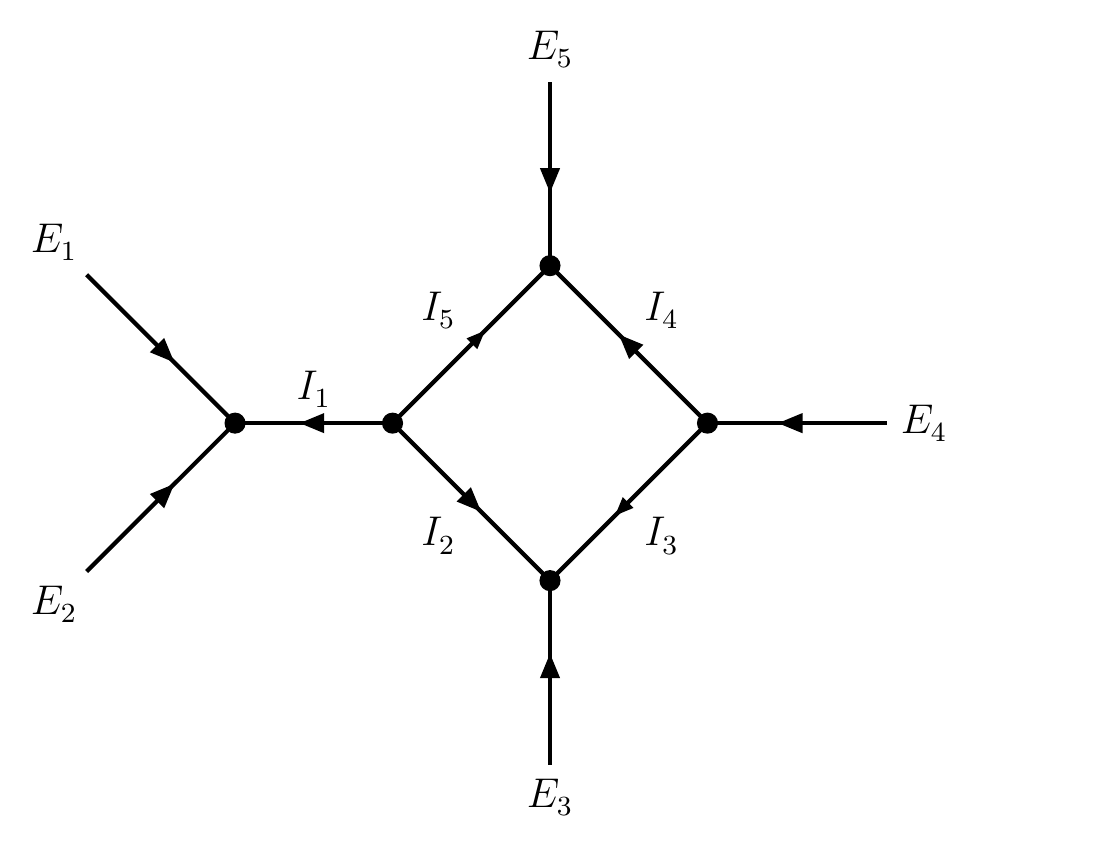}}\\\hline
Operator & Diagram & $E_{1}$ & $E_{2}$ & $E_{3}$ & $E_{4}$ & $E_{5}$ &
Operator & Diagram & $E_{1}$ & $E_{2}$ & $E_{3}$ & $E_{4}$ & $E_{5}$ \\\hline
\multirow{11}{*}{NL1} & \multirow{4}{*}{\begin{tabular}{@{}c@{}}NL-1-2-1\\NL-1-2-5\end{tabular}}& $\overline{\ell}$  & $\overline{\ell}$ & $\overline{Q}$ & $H^{\dagger}$ & $d_{R}$  &\multirow{5}{*}{NL2} & \multirow{2}{*}{\begin{tabular}{@{}c@{}}NL-1-2-1\\NL-1-2-5\end{tabular}}& $\overline{\ell}$  & $\overline{\ell}$ & $\bar{u}_{R}$ & $H^{\dagger}$ & $Q$\\ \cline{3-7}\cline{10-14}
& &  $\overline{\ell}$ & $\overline{Q}$ & $\overline{\ell}$  & $H^{\dagger}$ & $d_{R}$ &  & & $Q$ & $\bar{u}_{R}$ & $\overline{\ell}$ & $H^{\dagger}$ & $\overline{\ell}$ \\ \cline{3-7}\cline{9-14}
& & $d_{R}$  & $\overline{\ell}$ & $\overline{Q}$ & $H^{\dagger}$ & $\overline{\ell}$ &  & \multirow{3}{*}{\begin{tabular}{@{}c@{}}NL-1-2-2\\NL-1-2-4\end{tabular}} & $\overline{\ell}$ & $\overline{\ell}$ & $H^{\dagger}$ & $\bar{u}_{R}$ & $Q$  \\ \cline{3-7}\cline{10-14}
& & $d_{R}$ & $\overline{Q}$ & $\overline{\ell}$  & $H^{\dagger}$ & $\overline{\ell}$ &   &  & $\overline{\ell}$ & $\overline{\ell}$ & $H^{\dagger}$ & $Q$ & $\bar{u}_{R}$  \\ \cline{2-7}\cline{10-14}
&  \multirow{7}{*}{\begin{tabular}{@{}c@{}}NL-1-2-2\\NL-1-2-4\end{tabular}} & $\overline{Q}$ & $\overline{\ell}$ & $H^{\dagger}$ & $\overline{\ell}$   & $d_{R}$  & & & $Q$ & $\bar{u}_{R}$ & $H^{\dagger}$ & $\overline{\ell}$ & $\overline{\ell}$  \\ \cline{3-7}\cline{8-14}
&  & $d_{R}$ & $\overline{\ell}$ & $H^{\dagger}$ & $\overline{\ell}$  & $\overline{Q}$
& \multirow{6}{*}{NL3} & \multirow{2}{*}{\begin{tabular}{@{}c@{}}NL-1-2-1\\NL-1-2-5\end{tabular}}& $d_{R}$ & $\overline{\ell}$ & $\bar{u}_{R}$ & $H^{\dagger}$ & $\bar{e}_{R}$   \\ \cline{3-7}\cline{10-14}
& & $\overline{\ell}$ & $\overline{\ell}$ & $H^{\dagger}$  & $\overline{Q}$  & $d_{R}$ &  & & $\bar{e}_{R}$ & $\bar{u}_{R}$ & $\overline{\ell}$  & $H^{\dagger}$ & $d_{R}$  \\ \cline{3-7}\cline{9-14}
& &  $d_{R}$ & $\overline{\ell}$ & $H^{\dagger}$  & $\overline{Q}$ & $\overline{\ell}$ & & \multirow{4}{*}{\begin{tabular}{@{}c@{}}NL-1-2-2\\NL-1-2-4\end{tabular}}& $d_{R}$ & $\overline{\ell}$ & $H^{\dagger}$ & $\bar{e}_{R}$ & $\bar{u}_{R}$ \\ \cline{3-7}\cline{10-14}
& & $\overline{\ell}$ & $\overline{\ell}$ & $H^{\dagger}$ & $d_{R}$ & $\overline{Q}$  &  & & $d_{R}$ & $\overline{\ell}$ & $H^{\dagger}$ & $\bar{u}_{R}$ & $\bar{e}_{R}$ \\ \cline{3-7}\cline{10-14}
& & $\overline{\ell}$ & $\overline{Q}$  & $H^{\dagger}$  & $d_{R}$ & $\overline{\ell}$ &  & & $\bar{u}_{R}$ & $\bar{e}_{R}$ & $H^{\dagger}$ & $\overline{\ell}$ & $d_{R}$ \\ \cline{3-7}\cline{10-14}
& & $d_{R}$ & $\overline{Q}$ & $H^{\dagger}$ & $\overline{\ell}$ & $\overline{\ell}$  &  & & $\bar{e}_{R}$ & $\bar{u}_{R}$  & $H^{\dagger}$ & $d_{R}$ & $\overline{\ell}$ \\ \hline
\pagebreak
%\multicolumn{14}{|r|}{Continued on next page}\\
%\multicolumn{14}{|c|}{Continued from table~\ref{tab:NL-2ex}}\\ \hline
\multirow{12}{*}{NL1} & \multirow{12}{*}{NL-1-2-3} & $H^{\dagger}$ & $\overline{\ell}$ & $\overline{Q}$  & $\overline{\ell}$ & $d_{R}$  &
\multirow{24}{*}{NL3} & \multirow{24}{*}{NL-1-2-3} & $H^{\dagger}$ & $\bar{e}_{R}$ & $\bar{u}_{R}$  & $\overline{\ell}$ & $d_{R}$ \\
\cline{3-7}\cline{10-14}
& & $H^{\dagger}$ & $\overline{\ell}$ & $d_{R}$ & $\overline{\ell}$ & $\overline{Q}$ &  & & $H^{\dagger}$ & $\bar{e}_{R}$ & $d_{R}$ & $\overline{\ell}$ & $\bar{u}_{R}$ \\\cline{3-7}\cline{10-14}
& & $H^{\dagger}$ & $\overline{Q}$ & $\overline{\ell}$  & $\overline{\ell}$ & $d_{R}$ &  & & $H^{\dagger}$ & $\bar{u}_{R}$ & $\bar{e}_{R}$  & $\overline{\ell}$ & $d_{R}$ \\\cline{3-7}\cline{10-14}
& & $H^{\dagger}$ & $\overline{Q}$ & $d_{R}$ & $\overline{\ell}$  & $\overline{\ell}$ &  & & $H^{\dagger}$ & $\bar{u}_{R}$  & $d_{R}$ & $\overline{\ell}$ & $\bar{e}_{R}$ \\ \cline{3-7}\cline{10-14}
& & $H^{\dagger}$ & $d_{R}$ & $\overline{\ell}$  & $\overline{\ell}$ & $\overline{Q}$ &  & & $H^{\dagger}$ & $d_{R}$ & $\bar{e}_{R}$ & $\overline{\ell}$ & $\bar{u}_{R}$ \\ \cline{3-7}\cline{10-14}
& & $H^{\dagger}$ & $d_{R}$ & $\overline{Q}$ & $\overline{\ell}$ & $\overline{\ell}$  &  & & $H^{\dagger}$ & $d_{R}$ & $\bar{u}_{R}$  & $\overline{\ell}$ & $\bar{e}_{R}$ \\ \cline{3-7}\cline{10-14}
& & $H^{\dagger}$ & $\overline{\ell}$ & $\overline{\ell}$ & $\overline{Q}$  & $d_{R}$ &  & & $H^{\dagger}$ & $\overline{\ell}$ & $\bar{u}_{R}$ & $\bar{e}_{R}$ & $d_{R}$ \\ \cline{3-7}\cline{10-14}
& & $H^{\dagger}$ & $\overline{\ell}$ & $d_{R}$ & $\overline{Q}$ & $\overline{\ell}$ &  & & $H^{\dagger}$ & $\overline{\ell}$ & $d_{R}$ & $\bar{e}_{R}$ & $\bar{u}_{R}$  \\ \cline{3-7}\cline{10-14}
& & $H^{\dagger}$ & $d_{R}$ & $\overline{\ell}$ & $\overline{Q}$ & $\overline{\ell}$ &  & & $H^{\dagger}$ & $\bar{u}_{R}$  & $\overline{\ell}$ & $\bar{e}_{R}$ & $d_{R}$ \\ \cline{3-7}\cline{10-14}
& & $H^{\dagger}$ & $\overline{\ell}$ & $\overline{\ell}$ & $d_{R}$ & $\overline{Q}$ &  & & $H^{\dagger}$ & $\bar{u}_{R}$  & $d_{R}$ & $\bar{e}_{R}$ & $\overline{\ell}$ \\ \cline{3-7}\cline{10-14}
& & $H^{\dagger}$ & $\overline{\ell}$ & $\overline{Q}$ & $d_{R}$ & $\overline{\ell}$ &  & & $H^{\dagger}$ & $d_{R}$ & $\overline{\ell}$ & $\bar{e}_{R}$ & $\bar{u}_{R}$ \\ \cline{3-7}\cline{10-14}
& & $H^{\dagger}$ & $\overline{Q}$ & $\overline{\ell}$ & $d_{R}$ & $\overline{\ell}$ &  & & $H^{\dagger}$ & $d_{R}$ & $\bar{u}_{R}$ & $\bar{e}_{R}$ & $\overline{\ell}$ \\ \cline{1-7}\cline{10-14}
\multirow{12}{*}{NL2}& \multirow{12}{*}{NL-1-2-3} & $H^{\dagger}$ & $\overline{\ell}$ & $\bar{u}_{R}$ & $\overline{\ell}$ & $Q$ &  &
& $H^{\dagger}$ & $\overline{\ell}$ & $\bar{e}_{R}$ & $\bar{u}_{R}$  & $d_{R}$ \\ \cline{3-7}\cline{10-14}
& & $H^{\dagger}$ & $\overline{\ell}$ & $Q$ & $\overline{\ell}$ & $\bar{u}_{R}$ & & & $H^{\dagger}$ & $\overline{\ell}$ & $d_{R}$ & $\bar{u}_{R}$ & $\bar{e}_{R}$ \\ \cline{3-7}\cline{10-14}
& & $H^{\dagger}$ & $\bar{u}_{R}$ & $\overline{\ell}$ & $\overline{\ell}$ & $Q$ & & & $H^{\dagger}$ & $\bar{e}_{R}$ & $\overline{\ell}$ & $\bar{u}_{R}$  & $d_{R}$ \\ \cline{3-7}\cline{10-14}
& & $H^{\dagger}$ & $\bar{u}_{R}$ & $Q$ & $\overline{\ell}$ & $\overline{\ell}$ & & & $H^{\dagger}$ & $\bar{e}_{R}$ & $d_{R}$ & $\bar{u}_{R}$  & $\overline{\ell}$ \\ \cline{3-7}\cline{10-14}
& & $H^{\dagger}$ & $Q$ & $\overline{\ell}$ & $\overline{\ell}$ & $\bar{u}_{R}$ & & & $H^{\dagger}$ & $d_{R}$ & $\overline{\ell}$ & $\bar{u}_{R}$ & $\bar{e}_{R}$  \\ \cline{3-7}\cline{10-14}
& & $H^{\dagger}$ & $Q$ & $\bar{u}_{R}$ & $\overline{\ell}$ & $\overline{\ell}$ & & & $H^{\dagger}$ & $d_{R}$ & $\bar{e}_{R}$ & $\bar{u}_{R}$  & $\overline{\ell}$ \\ \cline{3-7}\cline{10-14}
& & $H^{\dagger}$ & $\overline{\ell}$ & $\overline{\ell}$ & $\bar{u}_{R}$ & $Q$ & & & $H^{\dagger}$ & $\overline{\ell}$ & $\bar{e}_{R}$ & $d_{R}$ & $\bar{u}_{R}$  \\ \cline{3-7}\cline{10-14}
& & $H^{\dagger}$ & $\overline{\ell}$ & $Q$ & $\bar{u}_{R}$ & $\overline{\ell}$ & & & $H^{\dagger}$ & $\overline{\ell}$ & $\bar{u}_{R}$  & $d_{R}$ & $\bar{e}_{R}$ \\ \cline{3-7}\cline{10-14}
& & $H^{\dagger}$ & $Q$ & $\overline{\ell}$ & $\bar{u}_{R}$ & $\overline{\ell}$ & & & $H^{\dagger}$ & $\bar{e}_{R}$ & $\overline{\ell}$ & $d_{R}$ & $\bar{u}_{R}$  \\ \cline{3-7}\cline{10-14}
& & $H^{\dagger}$ & $\overline{\ell}$ & $\overline{\ell}$ & $Q$ & $\bar{u}_{R}$ & & & $H^{\dagger}$ & $\bar{e}_{R}$ & $\bar{u}_{R}$ & $d_{R}$ & $\overline{\ell}$ \\ \cline{3-7}\cline{10-14}
& & $H^{\dagger}$ & $\overline{\ell}$ & $\bar{u}_{R}$ & $Q$ & $\overline{\ell}$ & & & $H^{\dagger}$ & $\bar{u}_{R}$  & $\overline{\ell}$ & $d_{R}$ & $\bar{e}_{R}$ \\ \cline{3-7}\cline{10-14}
& & $H^{\dagger}$ & $\bar{u}_{R}$ & $\overline{\ell}$ & $Q$ & $\overline{\ell}$ & & & $H^{\dagger}$ & $\bar{u}_{R}$ & $\bar{e}_{R}$ & $d_{R}$ & $\overline{\ell}$ \\\hline \hline
\caption{The possible external field attachments for the topology NL-1-2, where the external and internal fields are labelled as $E_i$ and $I_i$ respectively.}
\label{tab:NL-2ex}
\end{longtable}

\begin{table}[htbp]
\renewcommand{\tabcolsep}{0.3mm}
\renewcommand{\arraystretch}{1.3}
\centering
\begin{tabular}{|c|c|c|c|c|c|c||c|c|c|c|c|c|c|}
\hline\hline
\end{tabular}
\end{table}

%%%%%%%%%%%%%%%%

%NL-3
\begin{table}[htbp]
\renewcommand{\tabcolsep}{0.3mm}
\renewcommand{\arraystretch}{1.2}
\centering
\begin{tabular}{|c|c|c|c|c|c|c||c|c|c|c|c|c|c|}
\hline\hline
\multicolumn{14}{|c|}{\includegraphics[width=0.3\textwidth]{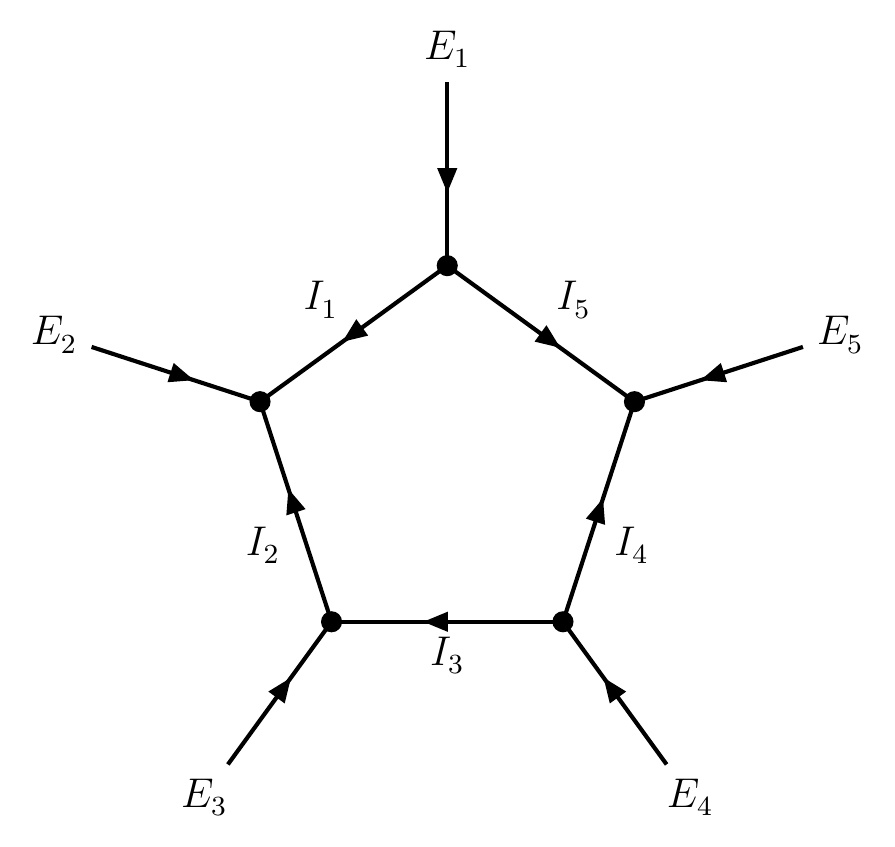}}\\
\hline
Operator & Diagram & $E_{1}$ & $E_{2}$ & $E_{3}$ & $E_{4}$ & $E_{5}$ &
Operator & Diagram & $E_{1}$ & $E_{2}$ & $E_{3}$ & $E_{4}$ & $E_{5}$ \\\hline
\multirow{6}{*}{NL1} & \multirow{6}{*}{\begin{tabular}{@{}c@{}}NL-1-3-1\\NL-1-3-2\end{tabular}}& $H^{\dagger}$ & $\overline{Q}$ & $\overline{\ell}$ & $d_{R}$ & $\overline{\ell}$  &
\multirow{12}{*}{NL3} & \multirow{12}{*}{\begin{tabular}{@{}c@{}}NL-1-3-1\\NL-1-3-2\end{tabular}}& $H^{\dagger}$ & $\bar{u}_{R}$ & $\bar{e}_{R}$ & $d_{R}$ & $\overline{\ell}$ \\ \cline{3-7}\cline{10-14}
& & $H^{\dagger}$ & $d_{R}$ & $\overline{\ell}$ & $\overline{Q}$ & $\overline{\ell}$ &  &  & $H^{\dagger}$ & $d_{R}$ & $\bar{e}_{R}$ & $\bar{u}_{R}$ & $\overline{\ell}$ \\ \cline{3-7}\cline{10-14}
& & $H^{\dagger}$ & $\overline{\ell}$ & $\overline{Q}$ & $d_{R}$ & $\overline{\ell}$ &  &  & $H^{\dagger}$ & $\bar{e}_{R}$ & $\bar{u}_{R}$ & $d_{R}$ & $\overline{\ell}$ \\ \cline{3-7}\cline{10-14}
& & $H^{\dagger}$ & $d_{R}$ & $\overline{Q}$ & $\overline{\ell}$ & $\overline{\ell}$ &  &  & $H^{\dagger}$ & $d_{R}$ & $\bar{u}_{R}$ & $\bar{e}_{R}$ & $\overline{\ell}$  \\ \cline{3-7}\cline{10-14}
& & $H^{\dagger}$ & $\overline{Q}$ & $d_{R}$ & $\overline{\ell}$ & $\overline{\ell}$ &  &  & $H^{\dagger}$ & $\bar{e}_{R}$ & $d_{R}$ & $\bar{u}_{R}$ & $\overline{\ell}$ \\ \cline{3-7}\cline{10-14}
& & $H^{\dagger}$ & $d_{R}$ & $\overline{\ell}$ & $\overline{\ell}$ & $\overline{Q}$ &  &  & $H^{\dagger}$ & $\bar{u}_{R}$ & $d_{R}$ & $\bar{e}_{R}$ & $\overline{\ell}$  \\\cline{1-7}\cline{10-14}
\multirow{6}{*}{NL2} & \multirow{6}{*}{\begin{tabular}{@{}c@{}}NL-1-3-1\\NL-1-3-2\end{tabular}}& $H^{\dagger}$ & $\bar{u}_{R}$ & $\overline{\ell}$ & $Q$ & $\overline{\ell}$  &  & & $H^{\dagger}$ & $\bar{u}_{R}$ & $\overline{\ell}$ & $d_{R}$ & $\bar{e}_{R}$ \\ \cline{3-7}\cline{10-14}
& & $H^{\dagger}$ & $Q$ & $\overline{\ell}$ & $\bar{u}_{R}$ & $\overline{\ell}$ &  &  & $H^{\dagger}$ & $d_{R}$ & $\overline{\ell}$ & $\bar{u}_{R}$ & $\bar{e}_{R}$ \\ \cline{3-7}\cline{10-14}
& & $H^{\dagger}$ & $\overline{\ell}$ & $\bar{u}_{R}$ & $Q$ & $\overline{\ell}$ &  &  & $H^{\dagger}$ & $d_{R}$ & $\bar{u}_{R}$ & $\overline{\ell}$ & $\bar{e}_{R}$ \\ \cline{3-7}\cline{10-14}
& & $H^{\dagger}$ & $Q$ & $\bar{u}_{R}$ & $\overline{\ell}$ & $\overline{\ell}$ &  &  & $H^{\dagger}$ & $\bar{u}_{R}$ & $d_{R}$ & $\overline{\ell}$ & $\bar{e}_{R}$ \\ \cline{3-7}\cline{10-14}
& & $H^{\dagger}$ & $\bar{u}_{R}$ & $Q$ & $\overline{\ell}$ & $\overline{\ell}$ &  &  & $H^{\dagger}$ & $d_{R}$ & $\overline{\ell}$ & $\bar{e}_{R}$ & $\bar{u}_{R}$ \\ \cline{3-7}\cline{10-14}
& & $H^{\dagger}$ & $Q$ & $\overline{\ell}$ & $\overline{\ell}$ & $\bar{u}_{R}$ &  &  & $H^{\dagger}$ & $d_{R}$ & $\bar{e}_{R}$ & $\overline{\ell}$ & $\bar{u}_{R}$  \\
\hline\hline
\end{tabular}
\caption{The possible external field attachments for the topology NL-1-3, where the external and internal fields are labelled as $E_i$ and $I_i$ respectively.}
\label{tab:NL-3ex}
\end{table}

\subsubsection{\label{subsubsec:U(1)}$U(1)_{Y}$ quantum number assignments }

For any given diagram with external field attachments specified, one can straightforwardly determine the hypercharges of the internal messenger fields from of the $U(1)_Y$ invariance at each vertex. Since a plenty of diagrams are involved, we would like to determine the $U(1)_Y$ quantum numbers at the topology level rather than at the diagram level. Notice that the hypercharge $Y$ of a field is related to its electric charge $Q$ via the Gell-Mann-Nishijima formula $Q=T_3+Y$, where $T_3$ is the third component of the weak isospin. For the topology NL-1-1, the equations of hypercharge conservation are given by
\begin{eqnarray}
\nonumber&&Y_{E_1}+Y_{I_1}-Y_{I_2}=0,~~~~ Y_{E_2}+Y_{I_2}+Y_{I_3}=0,\\
 &&Y_{E_5}-Y_{I_1}-Y_{I_3}-Y_{I_4}=0,~~~Y_{E_3}+Y_{E_4}+Y_{I_4}=0\,,
\end{eqnarray}
where the labels $E_i$ and $I_i$ represent the external and internal fields respectively, the diagram can be found in table~\ref{tab:NL-1ex}. The solution to the above equations leads to the following constraints on the hypercharge:
\begin{eqnarray}
Y_{I_1}=\alpha,~~~Y_{I_2}=Y_{E_1}+\alpha,~~Y_{I_3}=-Y_{E_1}-Y_{E_2}-\alpha,~~~Y_{I_4}=-Y_{E_3}-Y_{E_4}\,,
\end{eqnarray}
where $\alpha$ is an arbitrary real parameter and it parameterizes the hypercharge flow in the loop. A definite value of $\alpha$ should be taken in a concrete model. For the second one-loop topology NL-1-2 as shown in table~\ref{tab:NL-2ex}, conservation of hypercharge at each vertex implies
\begin{eqnarray}
        \nonumber&&Y_{E_1}+Y_{E_2}+Y_{I_1}=0,~~~~Y_{I_1}+Y_{I_2}+Y_{I_5}=0,~~~~ Y_{E_3}+Y_{I_2}+Y_{I_3}=0,\\
        &&Y_{E_4}-Y_{I_3}-Y_{I_4}=0,~~~Y_{E_5}+Y_{I_4}+Y_{I_5}=0\,.
\end{eqnarray}
The solution to the above system of equations leads to the following constraints on hypercharge:
\begin{eqnarray}
\nonumber&&Y_{I_1}=-Y_{E_1}-Y_{E_2},~~~Y_{I_2}=\alpha,~~~Y_{I_3}=-Y_{E_3}-\alpha,\\
&&Y_{I_4}=Y_{E_3}+Y_{E_4}+\alpha,~~~Y_{I_5}=-Y_{E_1}-Y_{E_2}-\alpha\,.
\end{eqnarray}
Similar to previous case, the hypercharge is not unambiguously fixed, and the arbitrariness is encoded in the real free parameter $\alpha$. For the last topology NL-1-3 with labels defined in table~\ref{tab:NL-3ex}, the gauge invariance under $U(1)_Y$ leads to the following constraints
\begin{eqnarray}
\nonumber&&Y_{E_1}-Y_{I_1}-Y_{I_5}=0,~~~~Y_{E_2}+Y_{I_1}+Y_{I_2}=0,~~~~ Y_{E_3}-Y_{I_2}+Y_{I_3}=0,\\
        &&Y_{E_4}-Y_{I_3}-Y_{I_4}=0,~~~Y_{E_5}+Y_{I_4}+Y_{I_5}=0\,,
\end{eqnarray}
and the solution is given by
\begin{eqnarray}
\nonumber&&Y_{I_1}=\alpha,~~~Y_{I_2}=-Y_{E_2}-\alpha,~~~Y_{I_3}=-Y_{E_2}-Y_{E_3}-\alpha,\\
&&Y_{I_4}=-Y_{E_1}-Y_{E_5}+\alpha,~~~Y_{I_5}=Y_{E_1}-\alpha\,,
\end{eqnarray}
We summarize the above results for the hypercharge values of the internal fields in table~\ref{tab:NL-U1Y}.

\begin{table}[t!]
\renewcommand{\tabcolsep}{0.3mm}
\renewcommand{\arraystretch}{1.2}
\centering
\begin{tabular}{|c|c|c|c|c|c|}
\hline\hline
Topology & $Y_{I_{1}}$ & $Y_{I_{2}}$ & $Y_{I_{3}}$ & $Y_{I_{4}}$ & $Y_{I_{5}}$ \\\hline
NL-1-1 & $\alpha$ & $Y_{E_1}+\alpha$ & $-Y_{E_1}-Y_{E_2}-\alpha$ & $-Y_{E_3}-Y_{E_4}$ &$\backslash$\\   \hline
NL-1-2 & $-Y_{E_1}-Y_{E_2}$ & $\alpha$ & $-Y_{E_3}-\alpha$ & $Y_{E_3}+Y_{E_4}+\alpha$ & $-Y_{E_1}-Y_{E_2}-\alpha$\\ \hline
NL-1-3 & $\alpha$ & $-Y_{E_2}-\alpha$ & $-Y_{E_2}-Y_{E_3}-\alpha$ & $-Y_{E_1}-Y_{E_5}+\alpha$ & $Y_{E_1}-\alpha$\\ \hline\hline
\end{tabular}
\caption{The hypercharge of each internal line for the three renormalizable topologies of long-range $0\nu\beta\beta$ decay, and the conventions for the hypercharge flows are shown in tables~\ref{tab:NL-1ex}, \ref{tab:NL-2ex} and \ref{tab:NL-3ex}.   }
\label{tab:NL-U1Y}
\end{table}

Once the assignment of external legs is specified for any given diagram, one can straightforwardly extract the hypercharges of the mediators by using the general results collected in table~\ref{tab:NL-U1Y}. Taking the diagram NL3-1-1-1-1 as an example, we have the external fields $E_1=\bar{u}_{R}$, $E_2=\bar{e}_{R}$, $E_3=\overline{\ell}$, $E_4=d_{R}$ and $E_5=H^{\dagger}$, thus the hypercharges of messenger fields are fixed to be
\begin{eqnarray}
Y_{S_1}=\alpha,~~~Y_{F_1}=-\frac{2}{3}+\alpha,~~~Y_{S_2}=-\frac{1}{3}-\alpha,~~~Y_{S_3}=-\frac{1}{6}\,,
\end{eqnarray}
where $Y_{u_R}=\frac{2}{3}$, $Y_{e_R}=-1$, $Y_{\ell}=-\frac{1}{2}$, $Y_{d_R}=-\frac{1}{3}$ and $Y_{H}=\frac{1}{2}$ have been used. Similarly, we have $E_1=\overline{\ell}$, $E_2=d_{R}$, $E_3=\bar{e}_{R}$, $E_4=\bar{u}_{R}$ and $E_5=H^{\dagger}$ for the diagram NL3-1-1-1-2 and consequently the hypercharge can be determined as follows,
\begin{eqnarray}
Y_{S_1}=\alpha,~~~Y_{F_1}=\frac{1}{2}+\alpha,~~~Y_{S_2}=-\frac{1}{6}-\alpha,~~Y_{S_3}=-\frac{1}{3}\,.
\end{eqnarray}

\subsubsection{\label{subsubsec:SU(2)}$SU(2)_{L}$ quantum number assignments}

Once the attachment of external legs to the fields of $0\nu\beta\beta$ decay operators is finished, as summarized in table~\ref{tab:NL-1ex}, table~\ref{tab:NL-2ex} and table~\ref{tab:NL-3ex}, the $SU(2)_{L}$ transformation of the each external line can be read off directly. We would like to mention that $\ell$, $Q$ and $H$ are $SU(2)_{L}$ doublets while $e_{R}$, $u_R$ and $d_R$ are $SU(2)_{L}$ singlets, and the complex conjugate of any $SU(2)_L$ irreducible representation is equivalent to itself. If focusing on the $SU(2)_{L}$ transformation of the external lines and ignoring other properties, from table~\ref{tab:NL-1ex} we can see that there are only 4 different $SU(2)_L$ assignments of external legs for the topology NL-1-1. Renormalizability fixes possible vertices to be only three and four point interactions. The trilinear couplings can be of the types fermion-fermion-scalar (FFS) or scalar-scalar-scalar (SSS), and the 4-point vertex can only be the scalar-scalar-scalar-scalar (SSSS) interaction. Accordingly, the interaction Lagrangian can be written as $\overline{F}_1F_2S$, $S_1S_2S_3$ and $S_1S_2S_3S_4$ respectively, the $SU(2)_L$ invariance gives the following constraints:
\begin{eqnarray}
\nonumber \overline{F}_1F_2S&:&~ n_{F_1}\otimes n_{F_2}\otimes n_S\supset \mathbf{1}\,,\\
\nonumber S_1S_2S_3&:&~ n_{S_1}\otimes n_{S_2}\otimes n_{S_3}\supset \mathbf{1}\,,\\
\label{eq:SU2-Gauge-Inv}S_1S_2S_3S_4 &:&~ n_{S_1}\otimes n_{S_2}\otimes n_{S_3}\otimes n_{S_4}\supset \mathbf{1}\,,
\end{eqnarray}
where $n_{X}$ denotes the $SU(2)_L$ representation under which the $X$ field transforms. The $SU(2)_{L}$ quantum number assignments for the internal fields can be determined by solving the constraint of Eq.~\eqref{eq:SU2-Gauge-Inv} at each interaction vertex. There are generally an infinite number of possible $SU(2)_L$ quantum numbers assignments to the internal particles except the tree-level diagrams. In the following, we will only consider singlet, doublet and triplet of $SU(2)_L$ for illustration, and extension to high dimensional representations is straightforward. We use the Mathematica group package \textbf{GroupMath}~\cite{Fonseca:2020vke} to efficiently determine the $SU(2)_L$ assignments. The results of $SU(2)_L$ quantum number assignments for the topology NL-1-1 are listed in table~\ref{tab:NL1-1SU2SU3}. The $SU(2)_{L}$ assignments for the other two topologies NL-1-2 and NL-1-3 can be determined in a similar way, and the results are listed in table~\ref{tab:NL1-2SU2SU3} and table~\ref{tab:NL1-3SU2SU3} respectively.

%NL-1
\begin{table}[htbp]
\renewcommand{\tabcolsep}{0.3mm}
\renewcommand{\arraystretch}{1.15}
\centering
\begin{tabular}{|c|c|c|c|c|c|c|c|c|c||c|c|c|c|c|c|c|c|c|}
\hline\hline
\multicolumn{19}{|c|}{\includegraphics[width=0.3\textwidth]{NL1-1.pdf}}\\
\hline
& $E_{1}$ & $E_{2}$ & $E_{3}$ & $E_{4}$ & $E_{5}$ & $I_{1}$ & $I_{2}$ & $I_{3}$ & $I_{4}$ & $E_{1}$ & $E_{2}$ & $E_{3}$ & $E_{4}$ & $E_{5}$ &
$I_{1}$ & $I_{2}$ & $I_{3}$ & $I_{4}$ \\\hline
\multirow{10}{*}{$SU(2)_L$} &\multirow{8}{*}{$2$} & \multirow{8}{*}{$1$} & \multirow{8}{*}{$2$} & \multirow{8}{*}{$2$} & \multirow{8}{*}{$2$} &
$2$ & $1$ & $1$ & $1$ & \multirow{3}{*}{2} & \multirow{3}{*}{2} & \multirow{3}{*}{2} & \multirow{3}{*}{1} & \multirow{3}{*}{2} &
$2$ & $1$ & $2$ & $2$ \\ \cline{7-10} \cline{16-19}
& & & & & & $2$ & $1$ & $1$ & $3$ &
& & & & & $2$ & $3$ & $2$ & $2$\\ \cline{7-10} \cline{16-19}
& & & & & & $1$ & $2$ & $2$ & $1$ & & & & & & $3$ & $2$ & $3$ & $2$\\ \cline{7-10} \cline{11-19}
& & & & & & $1$ & $2$ & $2$ & $3$ & \multirow{3}{*}{$1$} & \multirow{3}{*}{$1$} & \multirow{3}{*}{$2$} & \multirow{3}{*}{$1$} & \multirow{3}{*}{$2$} & $1$ & $1$ & $1$ & $2$\\ \cline{7-10} \cline{16-19}
& & & & & & $3$ & $2$ & $2$ & $1$ & & & & & & $2$ & $2$ & $2$ & $2$\\ \cline{7-10} \cline{16-19}
& & & & & & $3$ & $2$ & $2$ & $3$ & & & & & & $3$ & $3$ & $3$ & $2$\\ \cline{7-10} \cline{11-19}
& & & & & & $2$ & $3$ & $3$ & $1$ & \multirow{4}{*}{$2$} & \multirow{4}{*}{$1$} & \multirow{4}{*}{$1$} & \multirow{4}{*}{$1$} & \multirow{4}{*}{$2$} & $2$ & $1$ & $1$ & $1$\\ \cline{7-10} \cline{16-19}
& & & & & & $2$ & $3$ & $3$ & $3$ & & & & & & $1$ & $2$ & $2$ & $1$\\ \cline{2-10} \cline{16-19}
& \multirow{2}{*}{$2$} & \multirow{2}{*}{$2$} & \multirow{2}{*}{$2$} & \multirow{2}{*}{$1$} & \multirow{2}{*}{$2$} & $1$ & $2$ & $1$ & $2$ &
& & & & & $3$ & $2$ & $2$ & $1$\\ \cline{7-10} \cline{16-19}
& & & & & & $1$ & $2$ & $3$ & $2$ & & & & & & $2$ & $3$ & $3$ & $1$\\ \hline\hline
\multirow{18}{*}{$SU(3)_C$} &\multirow{9}{*}{$1$} & \multirow{9}{*}{$3$} & \multirow{9}{*}{$1$} & \multirow{9}{*}{$\bar{3}$} & \multirow{9}{*}{$1$} & $\bar{3}$ & $\bar{3}$ & $1$ & $3$ &
\multirow{9}{*}{$1$} & \multirow{9}{*}{$\bar{3}$} & \multirow{9}{*}{$1$} & \multirow{9}{*}{$3$} & \multirow{9}{*}{$1$} & $3$ & $3$ & $1$ & $\bar{3}$\\ \cline{7-10} \cline{16-19}
& & & & & & $3$ & $3$ & $3$ & $3$ & & & & & & $1$ & $1$ & $3$ & $\bar{3}$\\ \cline{7-10} \cline{16-19}
& & & & & & $\bar{6}$ & $\bar{6}$ & $3$ & $3$ & & & & & & $8$ & $8$ & $3$ & $\bar{3}$\\ \cline{7-10} \cline{16-19}
& & & & & & $1$ & $1$ & $\bar{3}$ & $3$ & & & & & & $\bar{3}$ & $\bar{3}$ & $\bar{3}$ & $\bar{3}$\\ \cline{7-10} \cline{16-19}
& & & & & & $8$ & $8$ & $\bar{3}$ & $3$ & & & & & & $6$ & $6$ & $\bar{3}$ & $\bar{3}$\\ \cline{7-10} \cline{16-19}
& & & & & & $3$ & $3$ & $\bar{6}$ & $3$ & & & & & & $8$ & $8$ & $\bar{6}$ & $\bar{3}$\\ \cline{7-10} \cline{16-19}
& & & & & & $8$ & $8$ & $6$ & $3$ &
& & & & & $\bar{3}$ & $\bar{3}$ & $6$ & $\bar{3}$\\ \cline{7-10} \cline{16-19}
& & & & & & $\bar{3}$ & $\bar{3}$ & $8$ & $3$ & & & & & & $3$ & $3$ & $8$ & $\bar{3}$\\ \cline{7-10} \cline{16-19}
& & & & & & $6$ & $6$ & $8$ & $3$ & & & & & & $\bar{6}$ & $\bar{6}$ & $8$ & $\bar{3}$\\ \cline{2-19}
&\multirow{9}{*}{$\bar{3}$} & \multirow{9}{*}{$3$} & \multirow{9}{*}{$1$} & \multirow{9}{*}{$1$} & \multirow{9}{*}{$1$} & $1$ & $\bar{3}$ & $1$ & $1$ &
\multirow{7}{*}{$1$} & \multirow{7}{*}{$1$} & \multirow{7}{*}{$\bar{3}$} & \multirow{7}{*}{$3$} & \multirow{7}{*}{$1$} & $1$ & $1$ & $1$ & $1$\\ \cline{7-10} \cline{16-19}
& & & & & & $\bar{3}$ & $3$ & $3$ & $1$ & & & & & & $\bar{3}$ & $\bar{3}$ & $3 $ & $1$\\ \cline{7-10} \cline{16-19}
& & & & & & $\bar{3}$ & $\bar{6}$ & $3$ & $1$ & & & & & & $\bar{3}$ & $\bar{3}$ & $3 $ & $8$\\ \cline{7-10} \cline{16-19}
& & & & & & $3$ & $1$ & $\bar{3}$ & $1$ &
& & & & & $6$ & $6$ & $\bar{6}$ & $1$\\ \cline{7-10} \cline{16-19}
& & & & & & $3$ & $8$ & $\bar{3}$ & $1$ &
& & & & & $6$ & $6$ & $\bar{6}$ & $8$\\ \cline{7-10} \cline{16-19}
& & & & & & $6$ & $3$ & $\bar{6}$ & $1$ & & & & & & $8$ & $8$ & $8$ & $1$\\ \cline{7-10} \cline{16-19}
& & & & & & $\bar{6}$ & $8$ & $6$ & $1$ & & & & & & $8$ & $8$ & $8$ & $8$\\ \cline{7-10} \cline{11-19}
& & & & & & $8$ & $\bar{3}$ & $8$ & $1$ & \multicolumn{9}{c|}{ }\\ \cline{7-10}
& & & & & & $8$ & $6$ & $8$ & $1$ & \multicolumn{9}{c|}{ }\\ \hline \hline
\end{tabular}
\caption{The independent $SU(2)_L$ and $SU(3)_C$ quantum number assignments for the topology NL-1-1, where $E_i$ and $I_i$ denote the external fields and internal fields respectively. Other assignments are related to these in the table through permutations of external and internal lines. }
\label{tab:NL1-1SU2SU3}
\end{table}
%NL-2 SU2
\clearpage
\renewcommand*{\tabcolsep}{0.5mm}
\renewcommand*{\arraystretch}{1.15}
\begin{longtable}{|c|c|c|c|c|c|c|c|c|c|c||c|c|c|c|c|c|c|c|c|c|}
\hline\hline
\endfirsthead

\hline
& $E_{1}$ & $E_{2}$ & $E_{3}$ & $E_{4}$ & $E_{5}$ & $I_{1}$ & $I_{2}$ & $I_{3}$ & $I_{4}$ & $I_{5}$ & $E_{1}$ & $E_{2}$ & $E_{3}$ & $E_{4}$ & $E_{5}$ & $I_{1}$ & $I_{2}$ & $I_{3}$ & $I_{4}$ & $I_{5}$\\ \hline
\endhead

\hline \multicolumn{21}{|r|}{Continued on next page}\\ \hline
\endfoot

\endlastfoot
\multicolumn{21}{|c|}{\includegraphics[width=0.4\textwidth]{NL1-2.pdf}}\\\hline
& $E_{1}$ & $E_{2}$ & $E_{3}$ & $E_{4}$ & $E_{5}$ &
$I_{1}$ & $I_{2}$ & $I_{3}$ & $I_{4}$ & $I_{5}$ &
$E_{1}$ & $E_{2}$ & $E_{3}$ & $E_{4}$ & $E_{5}$ &
$I_{1}$ & $I_{2}$ & $I_{3}$ & $I_{4}$ & $I_{5}$\\ \hline
\multirow{20}{*}{$SU(2)_L$} & \multirow{9}{*}{$2$} & \multirow{9}{*}{$2$} & \multirow{9}{*}{$2$} & \multirow{9}{*}{$2$} & \multirow{9}{*}{$1$} & $1$ & $1$ & $2$ & $1$ & $1$ & \multirow{8}{*}{$2$} & \multirow{8}{*}{$2$} & \multirow{8}{*}{$2$} & \multirow{8}{*}{$1$} & \multirow{8}{*}{$2$} & $1$ & $2$ & $2$ & $2$ & $1$ \\ \cline{7-11} \cline{17-21}
& &  &  &  &  & $3$ & $3$ & $2$ & $1$ & $1$ &  &  &  &  &  & $3$ & $1$ & $2$ & $2$ & $3$ \\ \cline{7-11} \cline{17-21}
& &  &  &  &  & $1$ & $2$ & $1$ & $2$ & $2$ &  &  &  &  &  & $1$ & $2$ & $1$ & $1$ & $2$ \\ \cline{7-11} \cline{17-21}
& &  &  &  &  & $3$ & $2$ & $1$ & $2$ & $2$ &  &  &  &  &  & $3$ & $2$ & $1$ & $1$ & $2$ \\ \cline{7-11} \cline{17-21}
& &  &  &  &  & $1$ & $2$ & $3$ & $2$ & $2$ &  &  &  &  &  & $1$ & $2$ & $3$ & $3$ & $2$ \\ \cline{7-11} \cline{17-21}
& &  &  &  &  & $3$ & $2$ & $3$ & $2$ & $2$ &  &  &  &  &  & $3$ & $2$ & $3$ & $3$ & $2$ \\ \cline{7-11} \cline{17-21}
& &  &  &  &  & $3$ & $1$ & $2$ & $3$ & $3$ &  &  &  &  &  & $1$ & $3$ & $2$ & $2$ & $3$ \\ \cline{7-11} \cline{17-21}
& &  &  &  &  & $1$ & $3$ & $2$ & $3$ & $3$ &  &  &  &  &  & $3$ & $3$ & $2$ & $2$ & $3$ \\ \cline{7-21}
& &  &  &  &  & $3$ & $3$ & $2$ & $3$ & $3$ & \multirow{4}{*}{$1$} & \multirow{4}{*}{$1$} & \multirow{4}{*}{$2$} & \multirow{4}{*}{$2$} & \multirow{4}{*}{$1$} & $1$ & $1$ & $2$ & $1$ & $1$ \\ \cline{2-11} \cline{17-21}
& \multirow{5}{*}{$2$} & \multirow{5}{*}{$2$} & \multirow{5}{*}{$1$} & \multirow{5}{*}{$1$} & \multirow{5}{*}{$1$} & $1$ & $1$ & $1$ & $1$ & $1$ &  &  &  &  &  & $1$ & $2$ & $1$ & $2$ & $2$ \\ \cline{7-11} \cline{17-21}
& &  &  &  &  & $1$ & $2$ & $2$ & $2$ & $2$ &  &  &  &  &  & $1$ & $2$ & $3$ & $2$ & $2$ \\ \cline{7-11} \cline{17-21}
& &  &  &  &  & $3$ & $2$ & $2$ & $2$ & $2$ &  &  &  &  &  & $1$ & $3$ & $2$ & $3$ & $3$ \\ \cline{7-21}
& &  &  &  &  & $1$ & $3$ & $3$ & $3$ & $3$ & \multirow{4}{*}{$1$} & \multirow{4}{*}{$2$} & \multirow{4}{*}{$2$} & \multirow{4}{*}{$1$} & \multirow{4}{*}{$1$} & $2$ & $1$ & $2$ & $2$ & $2$ \\ \cline{7-11} \cline{17-21}
& &  &  &  &  & $3$ & $3$ & $3$ & $3$ & $3$ &  &  &  &  &  & $2$ & $2$ & $1$ & $1$ & $1$ \\ \cline{2-11} \cline{17-21}
& \multirow{4}{*}{$1$} & \multirow{4}{*}{$2$} & \multirow{4}{*}{$2$} & \multirow{4}{*}{$2$} & \multirow{4}{*}{$2$} & $2$ & $1$ & $2$ & $1$ & $2$ &  &  &  &  &  & $2$ & $2$ & $3$ & $3$ & $3$ \\ \cline{7-11} \cline{17-21}
& &  &  &  &  & $2$ & $1$ & $2$ & $3$ & $2$ &  &  &  &  &  & $2$ & $3$ & $2$ & $2$ & $2$ \\ \cline{7-21}
 & &  &  &  &  & $2$ & $2$ & $1$ & $2$ & $3$ & \multirow{4}{*}{$1$} & \multirow{4}{*}{$1$} & \multirow{4}{*}{$2$} & \multirow{4}{*}{$1$} & \multirow{4}{*}{$2$} & $1$ & $1$ & $2$ & $2$ & $1$ \\ \cline{7-11} \cline{17-21}
& &  &  &  &  & $2$ & $2$ & $3$ & $2$ & $3$ &  &  &  &  &  & $1$ & $2$ & $1$ & $1$ & $2$ \\ \cline{2-11} \cline{17-21}
& \multirow{2}{*}{$1$} & \multirow{2}{*}{$2$} & \multirow{2}{*}{$1$} & \multirow{2}{*}{$2$} & \multirow{2}{*}{$1$} & $2$ & $1$ & $1$ & $2$ & $2$ &  &  &  &  &  & $1$ & $2$ & $3$ & $3$ & $2$ \\ \cline{7-11} \cline{17-21}
& &  &  &  &  & $2$ & $3$ & $3$ & $2$ & $2$ &  &  &  &  &  & $1$ & $3$ & $2$ & $2$ & $3$ \\\hline
% \multicolumn{21}{|r|}{Continued on next page}\\ \hline
\pagebreak
%\multicolumn{21}{|c|}{Continued from table~\ref{tab:NL1-2SU2SU3}}\\ \hline
\multirow{32}{*}{$SU(3)_C$}& \multirow{9}{*}{$1$} & \multirow{9}{*}{$1$} & \multirow{9}{*}{$\bar{3}$} & \multirow{9}{*}{$1$} & \multirow{9}{*}{$3$} & $1$ & $1$ & $3$ & $\bar{3}$ & $1$ & \multirow{9}{*}{$1$} & \multirow{9}{*}{$\bar{3}$} & \multirow{9}{*}{$1$} & \multirow{9}{*}{$1$} & \multirow{9}{*}{$3$} & $3$ & $\bar{3}$ & $3$ & $\bar{3}$ & $1$ \\ \cline{7-11} \cline{17-21}
& &  &  &  &  & $1$ & $\bar{3}$ & $\bar{3}$ & $3$ & $3$ &  &  &  &  &  & $3$ & $3$ & $\bar{3}$ & $3$ & $3$ \\ \cline{7-11} \cline{17-21}
& &  &  &  &  & $1$ & $\bar{3}$ & $6$ & $\bar{6}$ & $3$ &  &  &  &  &  & $3$ & $\bar{6}$ & $6$ & $\bar{6}$ & $3$ \\ \cline{7-11} \cline{17-21}
& &  &  &  &  & $1$ & $3$ & $1$ & $1$ & $\bar{3}$ &  &  &  &  &  & $3$ & $1$ & $1$ & $1$ & $\bar{3}$ \\ \cline{7-11} \cline{17-21}
& &  &  &  &  & $1$ & $3$ & $8$ & $8$ & $\bar{3}$ &  &  &  &  &  & $3$ & $8$ & $8$ & $8$ & $\bar{3}$ \\ \cline{7-11} \cline{17-21}
& &  &  &  &  & $1$ & $6$ & $\bar{3}$ & $3$ & $\bar{6}$ &  &  &  &  &  & $3$ & $3$ & $\bar{3}$ & $3$ & $\bar{6}$ \\ \cline{7-11} \cline{17-21}
& &  &  &  &  & $1$ & $\bar{6}$ & $8$ & $8$ & $6$ &  &  &  &  &  & $3$ & $8$ & $8$ & $8$ & $6$ \\ \cline{7-11} \cline{17-21}
& &  &  &  &  & $1$ & $8$ & $3$ & $\bar{3}$ & $8$ &  &  &  &  &  & $3$ & $\bar{3}$ & $3$ & $\bar{3}$ & $8$ \\ \cline{7-11} \cline{17-21}
& &  &  &  &  & $1$ & $8$ & $\bar{6}$ & $6$ & $8$ &  &  &  &  &  & $3$ & $6$ & $\bar{6}$ & $6$ & $8$ \\ \cline{2-21}
&\multirow{9}{*}{$3$} & \multirow{9}{*}{$1$} & \multirow{9}{*}{$\bar{3}$} & \multirow{9}{*}{$1$} & \multirow{9}{*}{$1$} & $\bar{3}$ & $1$ & $3$ & $\bar{3}$ & $3$ & \multirow{9}{*}{$1$} & \multirow{9}{*}{$1$} & \multirow{9}{*}{$3$} & \multirow{9}{*}{$\bar{3}$} & \multirow{9}{*}{$1$} & $1$ & $1$ & $\bar{3}$ & $1$ & $1$ \\ \cline{7-11} \cline{17-21}
& &  &  &  &  & $\bar{3}$ & $3$ & $1$ & $1$ & $1$ &  &  &  &  &  & $1$ & $\bar{3}$ & $1$ & $\bar{3}$ & $3$ \\ \cline{7-11} \cline{17-21}
& &  &  &  &  & $\bar{3}$ & $3$ & $8$ & $8$ & $8$ &  &  &  &  &  & $1$ & $\bar{3}$ & $8$ & $\bar{3}$ & $3$ \\ \cline{7-11} \cline{17-21}
& &  &  &  &  & $\bar{3}$ & $\bar{3}$ & $\bar{3}$ & $3$ & $\bar{3}$ &  &  &  &  &  & $1$ & $3$ & $3$ & $3$ & $\bar{3}$ \\ \cline{7-11} \cline{17-21}
& &  &  &  &  & $\bar{3}$ & $\bar{3}$ & $6$ & $\bar{6}$ & $6$ &  &  &  &  &  & $1$ & $3$ & $\bar{6}$ & $3$ & $\bar{3}$ \\ \cline{7-11} \cline{17-21}
& &  &  &  &  & $\bar{3}$ & $\bar{6}$ & $8$ & $8$ & $8$ &  &  &  &  &  & $1$ & $6$ & $8$ & $6$ & $\bar{6}$ \\ \cline{7-11} \cline{17-21}
& &  &  &  &  & $\bar{3}$ & $6$ & $\bar{3}$ & $3$ & $\bar{3}$ &  &  &  &  &  & $1$ & $\bar{6}$ & $3$ & $\bar{6}$ & $6$ \\ \cline{7-11} \cline{17-21}
& &  &  &  &  & $\bar{3}$ & $8$ & $3$ & $\bar{3}$ & $3$ &  &  &  &  &  & $1$ & $8$ & $\bar{3}$ & $8$ & $8$ \\ \cline{7-11} \cline{17-21}
& &  &  &  &  & $\bar{3}$ & $8$ & $\bar{6}$ & $6$ & $\bar{6}$ &  &  &  &  &  & $1$ & $8$ & $6$ & $8$ & $8$ \\ \cline{2-21}
&\multirow{5}{*}{$3$} & \multirow{5}{*}{$1$} & \multirow{5}{*}{$1$} & \multirow{5}{*}{$\bar{3}$} & \multirow{5}{*}{$1$} & $\bar{3}$ & $1$ & $1$ & $\bar{3}$ & $3$ & \multirow{5}{*}{$1$} & \multirow{5}{*}{$\bar{3}$} & \multirow{5}{*}{$1$} & \multirow{5}{*}{$3$} & \multirow{5}{*}{$1$} & $3$ & $1$ & $1$ & $3$ & $\bar{3}$ \\ \cline{7-11} \cline{17-21}
& &  &  &  &  & $\bar{3}$ & $3$ & $\bar{3}$ & $8$ & $8$ &  &  &  &  &  & $3$ & $3$ & $\bar{3}$ & $\bar{3}$ & $3$ \\ \cline{7-11} \cline{17-21}
& &  &  &  &  & $\bar{3}$ & $\bar{3}$ & $3$ & $3$ & $\bar{3}$ &  &  &  &  &  & $3$ & $3$ & $\bar{3}$ & $6$ & $\bar{6}$ \\ \cline{7-11} \cline{17-21}
& &  &  &  &  & $\bar{3}$ & $\bar{3}$ & $3$ & $\bar{6}$ & $6$ &  &  &  &  &  & $3$ & $\bar{3}$ & $3$ & $8$ & $8$ \\ \cline{7-11} \cline{17-21}
& &  &  &  &  & $\bar{3}$ & $\bar{6}$ & $6$ & $8$ & $8$ &  &  &  &  &  & $3$ & $6$ & $\bar{6}$ & $8$ & $8$ \\ \cline{2-21}
&\multirow{9}{*}{$1$} & \multirow{9}{*}{$1$} & \multirow{9}{*}{$1$} & \multirow{9}{*}{$3$} & \multirow{9}{*}{$\bar{3}$} & $1$ & $1$ & $1$ & $3$ & $1$ & \multirow{7}{*}{$3$} & \multirow{7}{*}{$\bar{3}$} & \multirow{7}{*}{$1$} & \multirow{7}{*}{$1$} & \multirow{7}{*}{$1$} & $1$ & $1$ & $1$ & $1$ & $1$ \\ \cline{7-11} \cline{17-21}
& &  &  &  &  & $1$ & $3$ & $\bar{3}$ & $\bar{3}$ & $\bar{3}$ &  &  &  &  &  & $1$ & $3$ & $\bar{3}$ & $3$ & $\bar{3}$ \\ \cline{7-11} \cline{17-21}
& &  &  &  &  & $1$ & $3$ & $\bar{3}$ & $6$ & $\bar{3}$ &  &  &  &  &  & $8$ & $3$ & $\bar{3}$ & $3$ & $\bar{3}$ \\ \cline{7-11} \cline{17-21}
& &  &  &  &  & $1$ & $\bar{3}$ & $3$ & $1$ & $3$ &  &  &  &  &  & $1$ & $\bar{6}$ & $6$ & $\bar{6}$ & $6$ \\ \cline{7-11} \cline{17-21}
& &  &  &  &  & $1$ & $\bar{3}$ & $3$ & $8$ & $3$ &  &  &  &  &  & $8$ & $\bar{6}$ & $6$ & $\bar{6}$ & $6$ \\ \cline{7-11} \cline{17-21}
& &  &  &  &  & $1$ & $\bar{6}$ & $6$ & $\bar{3}$ & $6$ &  &  &  &  &  & $1$ & $8$ & $8$ & $8$ & $8$ \\ \cline{7-11} \cline{17-21}
& &  &  &  &  & $1$ & $6$ & $\bar{6}$ & $8$ & $\bar{6}$ &  &  &  &  &  & $8$ & $8$ & $8$ & $8$ & $8$ \\ \cline{7-21}
& &  &  &  &  & $1$ & $8$ & $8$ & $3$ & $8$ & \multicolumn{10}{c|}{ }\\   \cline{7-11}
& &  &  &  &  & $1$ & $8$ & $8$ & $\bar{6}$ & $8$ &  \multicolumn{10}{c|}{ }\\ \hline\hline
\caption{The independent $SU(2)_L$ and $SU(3)_C$ quantum number assignments for the topology NL-1-2, where $E_i$ and $I_i$ denote the external fields and internal fields respectively. Other assignments are related to these in the table through permutations of external and internal lines.}
\label{tab:NL1-2SU2SU3}
\end{longtable}

%NL-3
\begin{table}[htbp]
\renewcommand{\tabcolsep}{0.5mm}
\renewcommand{\arraystretch}{1.3}
\centering
\begin{tabular}{|c|c|c|c|c|c|c|c|c|c|c||c|c|c|c|c|c|c|c|c|c|}\hline\hline
\multicolumn{21}{|c|}{\includegraphics[width=0.4\textwidth]{NL1-3.pdf}}\\\hline
& $E_{1}$ & $E_{2}$ & $E_{3}$ & $E_{4}$ & $E_{5}$ & $I_{1}$ & $I_{2}$ & $I_{3}$ & $I_{4}$ & $I_{5}$ & $E_{1}$ & $E_{2}$ & $E_{3}$ & $E_{4}$ & $E_{5}$ & $I_{1}$ & $I_{2}$ & $I_{3}$ & $I_{4}$ & $I_{5}$\\ \hline
\multirow{8}{*}{$SU(2)_L$} &\multirow{7}{*}{$2$} & \multirow{7}{*}{$2$} & \multirow{7}{*}{$2$} & \multirow{7}{*}{$2$} & \multirow{7}{*}{$1$} & $2$ & $1$ & $2$ & $1$ & $1$ & \multirow{4}{*}{$2$} & \multirow{4}{*}{$2$} & \multirow{4}{*}{$1$} & \multirow{4}{*}{$1$} & \multirow{4}{*}{$1$} &
$2$ & $1$ & $1$ & $1$ & $1$ \\ \cline{7-11} \cline{17-21}
& & & & & & $2$ & $1$ & $2$ & $3$ & $3$ & & & & & & $1$ & $2$ & $2$ & $2$ & $2$\\ \cline{7-11} \cline{17-21}
& & & & & & $1$ & $2$ & $1$ & $2$ & $2$ & & & & & & $3$ & $2$ & $2$ & $2$ & $2$\\ \cline{7-11} \cline{17-21}
& & & & & & $3$ & $2$ & $1$ & $2$ & $2$ & & & & & & $2$ & $3$ & $3$ & $3$ & $3$\\ \cline{7-11} \cline{12-21}
& & & & & & $3$ & $2$ & $3$ & $2$ & $2$ & \multirow{4}{*}{$1$} & \multirow{4}{*}{$2$} & \multirow{4}{*}{$1$} & \multirow{4}{*}{$2$} & \multirow{4}{*}{$1$} & $2$ & $1$ & $1$ & $2$ & $2$\\ \cline{7-11} \cline{17-21}
& & & & & & $2$ & $3$ & $2$ & $1$ & $1$ & & & & & & $1$ & $2$ & $2$ & $1$ & $1$\\ \cline{7-11} \cline{17-21}
& & & & & & $2$ & $3$ & $2$ & $3$ & $3$ & & & & & & $3$ & $2$ & $2$ & $3$ & $3$\\ \cline{2-11} \cline{17-21}
& \multicolumn{10}{c||}{ } & & & & & & $2$ & $3$ & $3$ & $2$ & $2$\\ \hline
\multirow{9}{*}{$SU(3)_C$} & \multirow{9}{*}{$1$} & \multirow{9}{*}{$1$} & \multirow{9}{*}{$\bar{3}$} & \multirow{9}{*}{$1$} & \multirow{9}{*}{$3$} & $1$ & $1$ & $3$ & $\bar{3}$ & $1$ & \multirow{9}{*}{$1$} & \multirow{9}{*}{$1$} & \multirow{9}{*}{$1$} & \multirow{9}{*}{$\bar{3}$ } & \multirow{9}{*}{$3$} & $1$ & $1$ & $1$ & $\bar{3}$ & $1$ \\ \cline{7-11} \cline{17-21}
& &  &  &  &  & $\bar{3}$ & $3$ & $\bar{3}$ & $3$ & $3$ &  &  &  &  &  & $\bar{3}$ & $3$ & $3$ & $3$ & $3$ \\ \cline{7-11} \cline{17-21}
& &  &  &  &  & $\bar{3}$ & $3$ & $6$ & $\bar{6}$ & $3$ &  &  &  &  &  & $\bar{3}$ & $3$ & $3$ & $\bar{6}$ & $3$ \\ \cline{7-11} \cline{17-21}
& &  &  &  &  & $3$ & $\bar{3}$ & $1$ & $1$ & $\bar{3}$ &  &  &  &  &  & $3$ & $\bar{3}$ & $\bar{3}$ & $1$ & $\bar{3}$ \\ \cline{7-11} \cline{17-21}
& &  &  &  &  & $3$ & $\bar{3}$ & $8$ & $8$ & $\bar{3}$ &  &  &  &  &  & $3$ & $\bar{3}$ & $\bar{3}$ & $8$ & $\bar{3}$ \\ \cline{7-11} \cline{17-21}
& &  &  &  &  & $6$ & $\bar{6}$ & $\bar{3}$ & $3$ & $\bar{6}$ &  &  &  &  &  & $6$ & $\bar{6}$ & $\bar{6}$ & $3$ & $\bar{6}$ \\ \cline{7-11} \cline{17-21}
& &  &  &  &  & $\bar{6}$ & $6$ & $8$ & $8$ & $6$ &  &  &  &  &  & $\bar{6}$ & $6$ & $6$ & $8$ & $6$ \\ \cline{7-11} \cline{17-21}
& &  &  &  &  & $8$ & $8$ & $3$ & $\bar{3}$ & $8$ &  &  &  &  &  & $8$ & $8$ & $8$ & $\bar{3}$ & $8$ \\ \cline{7-11} \cline{17-21}
& &  &  &  &  & $8$ & $8$ & $\bar{6}$ & $6$ & $8$ &  &  &  &  &  & $8$ & $8$ & $8$ & $6$ & $8$ \\ \hline\hline
\end{tabular}
\caption{The independent $SU(2)_L$ and $SU(3)_C$ quantum number assignments for the topology NL-1-3, where $E_i$ and $I_i$ denote the external fields and internal fields respectively. Other assignments are related to these in the table through permutations of external and internal lines. }
\label{tab:NL1-3SU2SU3}
\end{table}

\subsubsection{\label{subsubsec:SU(3)}$SU(3)_{C}$ quantum number assignments }

All the long-range $0\nu\beta\beta$ decay operators involve one quark, one anti-quark, two lepton fields and one Higgs field. The quark is in the irreducible representations 3, and the anti-quark is in the conjugate triplet representation $\bar{3}$ of $SU(3)_C$ while leptons and Higgs are invariant under the $SU(3)_C$ group. Hence two external fields transform as $3$ and $\bar{3}$ and the remaining three external fields are trivial singlets of $SU(3)_C$. It is remarkable that one can assign $SU(3)_C$ quantum numbers of external legs at the topology level without specifying the field property of each line. As regard the topology NL-1-1, the external field $E_5$ is attached to a four-point vertex, consequently it can only be the Higgs scalar field. We can see there are only four independent $SU(3)_C$ assignments to the external legs, without loss of generality we can choose $(E_1, E_2, E_3, E_4, E_5)\sim(1, 3, 1, \bar{3}, 1)$, $(\bar{3}, 3, 1, 1, 1)$, $(1, \bar{3}, 1, 3, 1)$ and $(1, 1, \bar{3}, 3, 1)$, as shown in table~\ref{tab:NL1-1SU2SU3}. Other assignments are redundant and they are related to these four representative ones by permutating external fields. For instance, the assignment $(E_1, E_2, E_3, E_4, E_5)\sim(1, 3, \bar{3}, 1, 1)$ is equivalent to $(E_1, E_2, E_3, E_4, E_5)\sim(1, 3, 1, \bar{3}, 1)$ since there is a $E_3, E_4$ permutation symmetry in the topology. Similarly there are eight different $SU(3)_C$ assignments to the external fields of the topology NL-1-2, as listed in table~\ref{tab:NL1-2SU2SU3}. Regarding the last topology NL-1-3, the five external lines attach at the vertices of a pentagon, and the two colored ones can be adjacent or spaced. Consequently it is sufficient to consider only two kinds of $SU(3)_C$ assignments displayed in table~\ref{tab:NL1-3SU2SU3}. Then we proceed to assign $SU(3)_C$ quantum numbers to each internal line. Similar to Eq.~\eqref{eq:SU2-Gauge-Inv} for $SU(2)_L$, one should determine whether $SU(3)_C$ invariant contractions can be formed at the vertices by using the technique of Young diagrams~\cite{Workman:2022ynf}, and this task can be made much easier with the help of the Mathematica package \texttt{GroupMath}~\cite{Fonseca:2020vke}. Analogous to the case of $SU(2)_L$ and $U(1)_Y$, there are in principle endless $SU(3)_C$ representation assignments consistent with the SM gauge invariance at one-loop level. We present the $SU(3)_C$ quantum numbers of the internal fields for the three topologies NL-1-1, NL-1-2 and NL-1-3 in table~\ref{tab:NL1-1SU2SU3}, table~\ref{tab:NL1-2SU2SU3} and table~\ref{tab:NL1-3SU2SU3} respectively, where only the lower-dimensional  $SU(3)_C$ representations $1$, $3$, $\bar{3}$, $6$, $\bar{6}$ and $8$ are used.

\subsubsection{\label{sec:Gmodelway}Constructing long-range $0\nu\beta\beta$ decay models }

Using the results of sections~\ref{subsubsec:U(1)}, \ref{subsubsec:SU(2)} and~\ref{subsubsec:SU(3)}, one can construct explicit UV models for $0\nu\beta\beta$ decay by assigning the SM $SU(3)_C\times SU(2)_L\times U(1)_Y$ quantum numbers to the internal fields. The first step is to choose a diagram, here we take NL-1-1-1 as an example. The second step is to attach fields to the external legs, see table~\ref{tab:NL-1ex} for different possibilities. We choose the first kind of attachment for the operator NL3, and it yields the diagrams shown in the left panel of figure~\ref{fig:external-NL3}. The third step is to determine the $U(1)_{Y}$ charges of the messenger fields by using table~\ref{tab:NL-U1Y}. The $U(1)_{Y}$ charges of external fields read as
\begin{align}
Y_{E_{1}}=Y_{\overline{u}_{R}}=-\frac{2}{3}\,,
Y_{E_{2}}=Y_{\overline{e}_{R}}=1\,,
Y_{E_{3}}=Y_{\overline{\ell}}=\frac{1}{2}\,,
Y_{E_{4}}=Y_{d_{R}}=-\frac{1}{3}\,,
Y_{E_{5}}=Y_{H^{\dagger}}=-\frac{1}{2}\,.
\end{align}
As a consequence, the $U(1)_{Y}$ charges of the internal fields are determined to be
\begin{align}
Y_{I_{1}}=\alpha\,,~~~Y_{I_{2}}=-\frac{2}{3}+\alpha\,,~~~
Y_{I_{3}}=-\frac{1}{3}-\alpha\,,~~~Y_{I_{4}}=\frac{5}{6}\,.
\end{align}

\begin{figure}[htbp]
\centering
\includegraphics[width=0.95\textwidth]{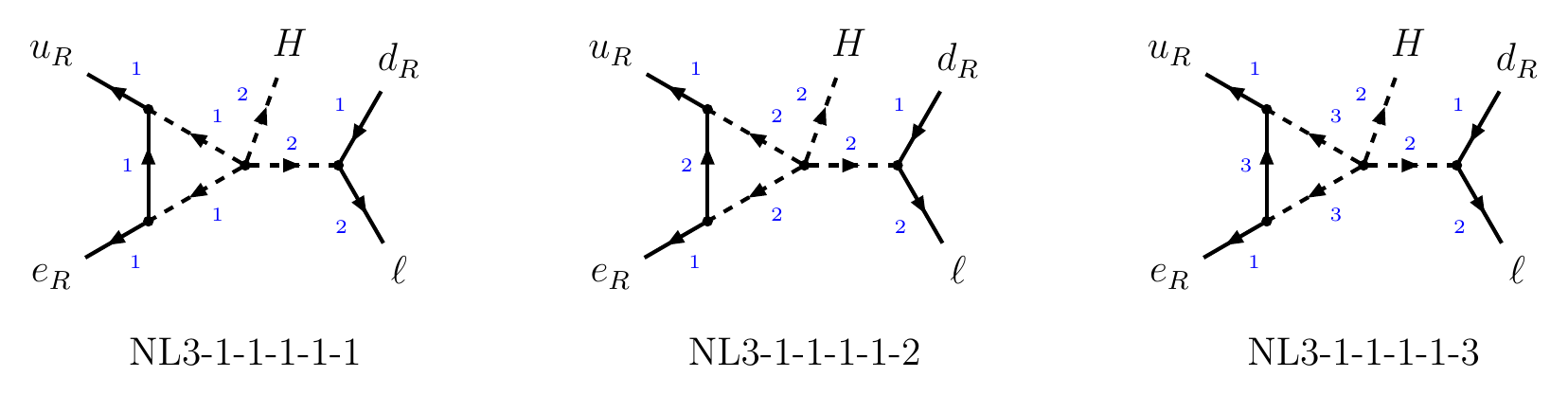}
\caption{Assignments of the $SU(2)_{L}$ quantum numbers for the diagram NL3-1-1-1-1.}
\label{fig:SU2-NL3}
\end{figure}
\begin{figure}[!htbp]
\centering
\includegraphics[width=0.99\textwidth]{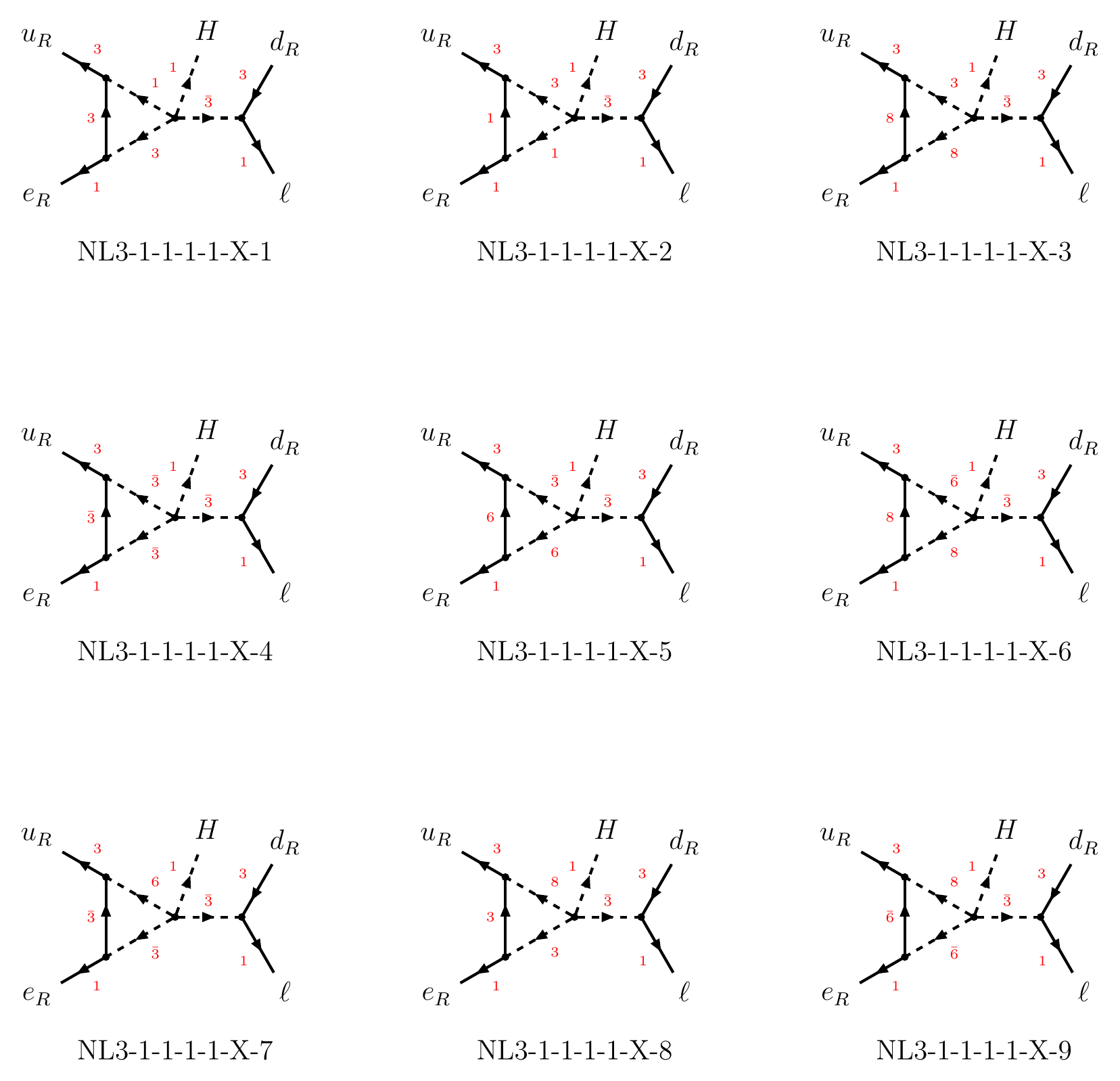}
\caption{Assignments of the $SU(3)_{C}$ quantum numbers for the diagram NL3-1-1-1-1-X, where X=1, 2, 3 stands for the possible $SU(2)_L$ quantum number assignments shown in figure~\ref{fig:SU2-NL3}.}
\label{fig:SU3-NL3}
\end{figure}

The fourth step is the assignment of the $SU(2)_L$ quantum numbers. The external fields transform as $(E_{1}, E_{2}, E_{3}, E_{4}, E_{5})\sim(1, 1, 2, 1, 2)$ under $SU(2)_L$. From table~\ref{tab:NL1-1SU2SU3} we see that the $SU(2)_{L}$ transformation of the mediators can be $(I_1, I_2, I_3, I_4)\sim(1, 1, 1, 2)$, $(2, 2, 2, 2)$, $(3, 3, 3, 2)$, these three $SU(2)_L$ assignments are displayed in figure~\ref{fig:SU2-NL3}. The last step is to determine the $SU(3)_{C}$ quantum numbers of messenger fields. The $SU(3)_C$ transformations of external legs are $(E_{1},E_{2},E_{3},E_{4},E_{5})=(\bar{3}, 1, 1, 3, 1)$. There is no such assignment in table~\ref{tab:NL1-1SU2SU3} at first glance. However, one can exchange $E_{1}$ and $E_{2}$ as well as $I_1$ and $I_3$ at topology level, consequently we can consider the assignment $(E_{1}, E_{2}, E_{3}, E_{4}, E_{5})\sim (1, \bar{3}, 1, 3, 1)$ instead. Then we see from table~\ref{tab:NL1-1SU2SU3} that the internal fields can transform as $(I_1, I_2, I_3, I_4)\sim(1, 3, 3, \bar{3})$, $(3, 1, 1, \bar{3})$, $(3, 8, 8, \bar{3})$, $(\bar{3}, \bar{3}, \bar{3}, \bar{3})$, $(\bar{3}, 6, 6, \bar{3})$, $(\bar{6}, 8, 8, \bar{3})$, $(6, \bar{3}, \bar{3}, \bar{3})$, $(8, 3, 3, \bar{3})$ and $(8, \bar{6}, \bar{6}, \bar{3})$ under $SU(3)_C$, as shown in figure~\ref{fig:SU3-NL3}. In this way, we can find the possible UV completions for all the long-range $0\nu\beta\beta$ decay operators.

\subsubsection{\label{sec:DetermineGe}Genuine one-loop models}

\begin{figure}[htbp]
\centering
\includegraphics[width=\textwidth]{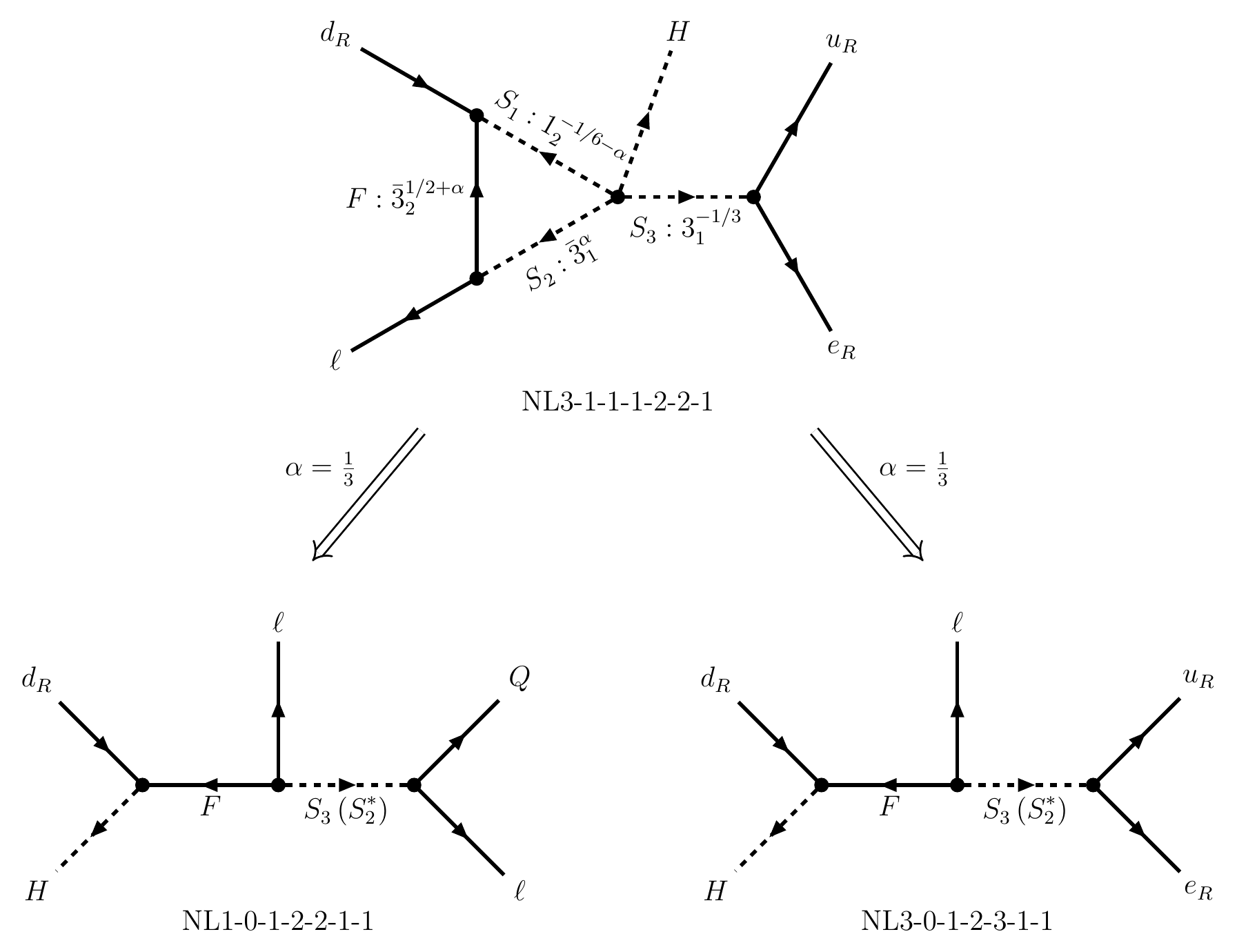}
\caption{One example of the non-genuine model NL3-1-1-1-2-2-1, and the messenger fields  $F$, $S_{2}$ and $S_{3}$ can lead to the tree-level contributions NL1-0-1-2-2-1-1 and NL3-0-1-2-3-1-1 in the case of $\alpha=\frac{1}{3}$. Notice that the mediators $F$, $S_{2}$ and $S_{3}$ will also lead to tree-level contributions NL1-0-1-2-2-1-1, NL3-0-1-2-3-1-1, NL3-0-1-1-3-1-1 when the hypercharge parameter $\alpha=\pm1$. The quantum numbers are given in the notation $C_{L}^{Y}$, where $C$ refers to the $SU(3)_{C}$ transformation, $L$ refers to the $SU(2)_{L}$ transformation, and $Y$ stands for $U(1)_{Y}$ charge. }
        \label{fig:genuine}
\end{figure}

Numerous one-loop models for long-range $0\nu\beta\beta$ decays can be generated through a series of steps described in previous sections, however, some of them are not the leading-order contribution to the long-range $0\nu\beta\beta$ decays. A one-loop model is the dominant contribution if and only if the combination of fields participating in the model can not generate
more important tree-level contributions to $0\nu\beta\beta$ decay. Such kind of models would be called genuine models, for which the tree-level diagrams are automatically absent without the need of invoking additional symmetries. If the lower order contributions can not be forbidden without extra symmetry, the model would be non-genuine. We can determine the genuineness of each model by comparing its field content with that of the tree-level models one by one. Since the quantum numbers of the mediators of the tree-level $0\nu\beta\beta$ decay models are unambiguously fixed, genuineness of a one-loop model generally excludes certain value of the hypercharge parameter $\alpha$.

We take the model NL3-1-1-1-2-2-1 for illustration, the Feynman diagram is shown in figure~\ref{fig:genuine}, in which we have introduced the notation $C_{L}^{Y}$ to label the quantum numbers of a field, where $C$ refers to the $SU(3)_{C}$ representation, $L$ refers to the $SU(2)_{L}$ transformation, and $Y$ stands for $U(1)_{Y}$ charge. When the hypercharge parameter $\alpha=\frac{1}{3}$, we see that the mediators $S_2$($S_3$) and $F$ as well as the associated interactions allow to generate the tree-level models NL1-0-1-2-2-1-1, NL3-0-1-2-3-1-1 which are the more important ones. Therefore the genuineness of the one-loop model NL3-1-1-1-2-2-1 requires $\alpha\neq\frac{1}{3}$. The condition of genuineness has been considered for each possible one-loop decomposition of the $0\nu\beta\beta$ decay operators, and the full results are listed in the attachment~\cite{Chen:2021sup2}.

\section{\label{sec:relaton to mass}Neutrino mass in long-range $0\nu\beta\beta$ decay models}

\begin{figure}[htbp]
\centering
\includegraphics[width=0.99\textwidth]{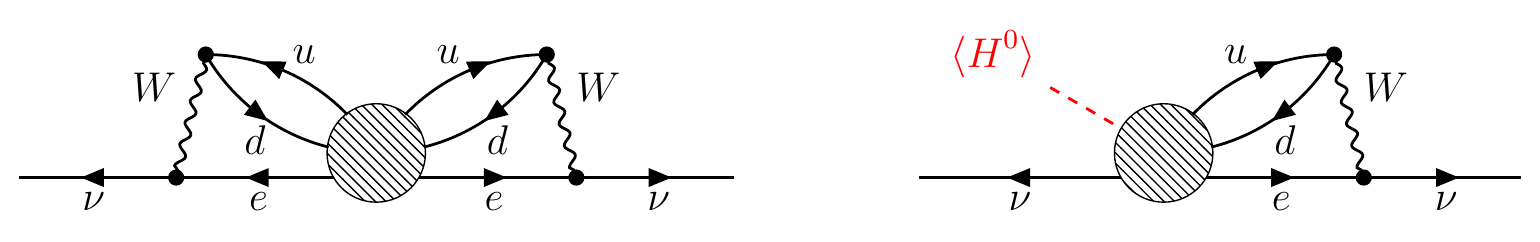}
\caption{The black box diagrams for neutrino masses from $0\nu\beta\beta$ decay effective vertices. The diagram in the left panel can generate Majorana neutrino masses if $0\nu\beta\beta$ decay is observed, while the diagram in the right panel can generate Majorana neutrino masses from the long-range $0\nu\beta\beta$ decay operator. In the UV decomposition of the current work, the effective vertex of the long-range $0\nu\beta\beta$ operator is realized by a one-loop diagram, which means the neutrino masses are generated at most at three-loop level. }
\label{fig:black_box}
\end{figure}

\begin{figure}[htbp]
\centering
\includegraphics[width=0.99\textwidth]{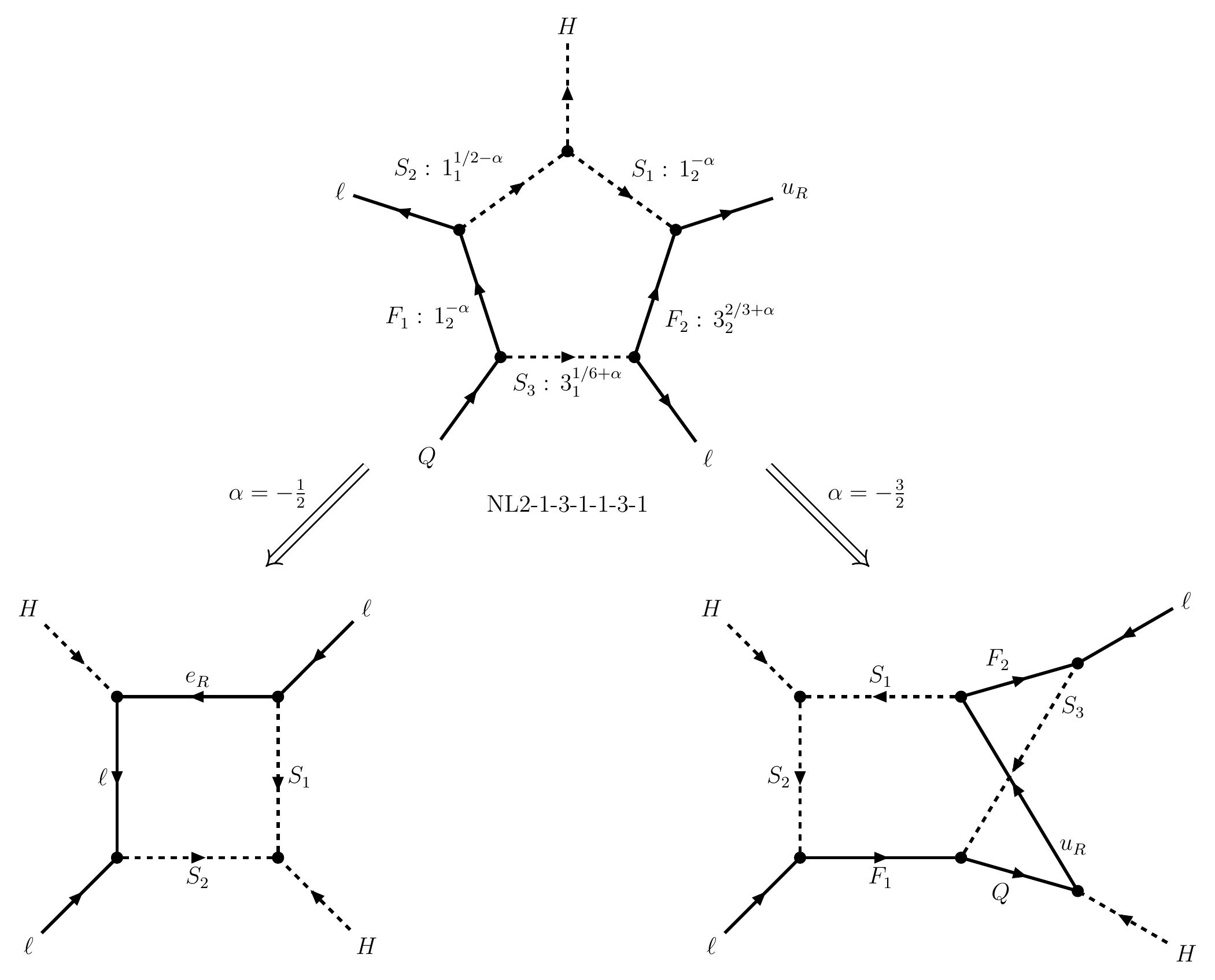}
\caption{The neutrino mass generation in the $0\nu\beta\beta$ decay model NL2-1-3-1-1-3-1. One sees that the mechanism producing neutrino mass depends on the value of the hypercharge parameter $\alpha$.}
\label{fig:RelatedToMass}
\end{figure}

The black box theorem shows that, one can obtain the Majorana neutrino masses by connecting the quark and charged lepton legs in these $0\nu\beta\beta$ decay effective operators with the SM interactions~\cite{Schechter:1981bd}, the schematic black box diagram is shown in left panel of figure~\ref{fig:black_box}. Since the long-range $0\nu\beta\beta$ decay operators violate lepton number by two units, we can similarly get another black box diagram for Majorana neutrino mass from the long-range $0\nu\beta\beta$ effective vertex, which is shown in the right panel of figure~\ref{fig:black_box}. Consequently, any $0\nu\beta\beta$ decay model will always generate a non-zero Majorana neutrino mass. In current work, the effective $0\nu\beta\beta$ decay operator in the black box diagram is realized at the one-loop level. The tree-level contribution is forbidden in order to maintain the genuineness of the one-loop model, so the black box is realized at most at the three-loop level in our UV models. The fields introduced in the one-loop $0\nu\beta\beta$ decay model can also generate neutrino mass. In some cases, these fields can result in a lower loop-level model for neutrino mass, then the three-loop diagram in the black box is the higher order contribution. In other words, one can construct neutrino mass diagrams by using the SM fields and the mediators that appear in one-loop renormalizable long-range $0\nu\beta\beta$ decay models at most at three-loop level. Indeed, as shown below, any decomposition of the long-range $0\nu\beta\beta$ decay operators contains automatically the particle content and interactions such that Majorana neutrino masses can be generated. Given the quantum numbers of mediators and the SM fields, one can use the Mathematica package \texttt{Sym2Int}~\cite{Fonseca:2017lem,Fonseca:2019yya} to generate all renormalizable interactions consistent with SM gauge symmetry. Subsequently we import these interactions to the package \texttt{qgraf}~\cite{Nogueira:1991ex} to generate all possible leading-order neutrino mass diagrams. We take the one-loop model NL2-1-3-1-1-3-1 for example, the leading order contribution to neutrino masses arises at one-loop and two-loop level for $\alpha=-1/2$ and $\alpha=-2/3$ respectively, as shown in figure~\ref{fig:RelatedToMass}. Since the black box theorem implies that any contributions to the $0\nu\beta\beta$ decay always induce Majorana neutrino masses, thus the contribution of the mass mechanism always exists in any $0\nu\beta\beta$ model. For models in which neutrino masses are generated at tree or one-loop level, one generally expects that the one-loop long-range contribution is subdominant to the mass mechanism, if the values of the model parameters are not severely fine-tuned. The long-range contribution and the mass mechanism can be comparable in certain parameter space for the one-loop $0\nu\beta\beta$ decay decomposition with two-loop or three-loop neutrino masses. Hence both tree and one-loop contributions to neutrino mass should be forbidden in a genuine one-loop model of long-range $0\nu\beta\beta$ decay, thus certain values of the hypercharge would be excluded.

\section{\label{sec:example}An example model of one-loop $0\nu\beta\beta$ decay }

\begin{figure}[htbp]
\centering
\includegraphics[width=\textwidth]{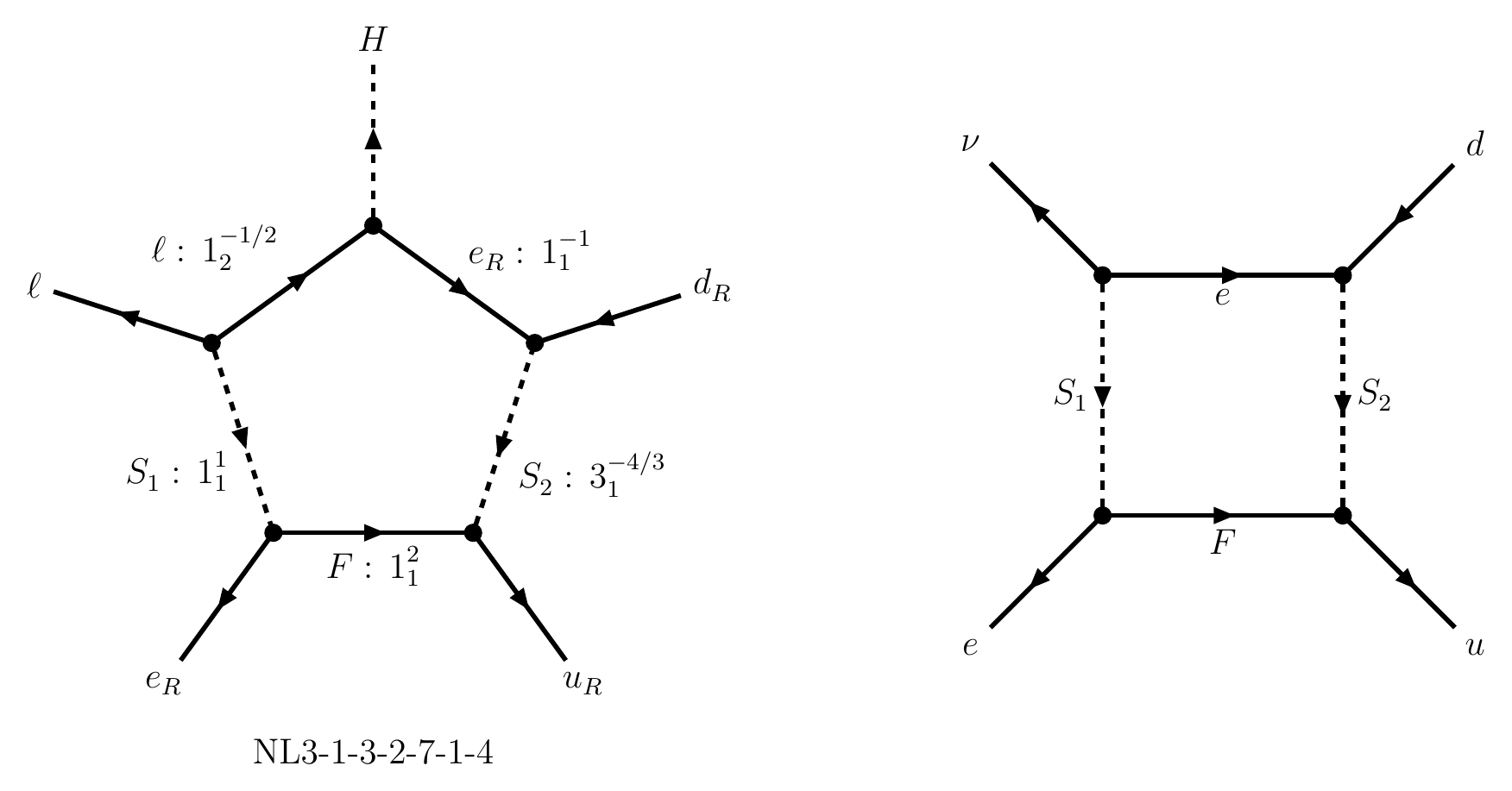}
\caption{The example model NL3-1-3-2-7-1-4 of one-loop $0\nu\beta\beta$ decay. After electroweak symmetry breaking, the corresponding Feynman diagram is shown in the right panel. For this model, the neutrino mass is generated at three-loop level, as shown in figure~\ref{fig:MassModelExample} and figure~\ref{fig:MassModelExample2}. }
\label{fig:example}
\end{figure}

In this section, we shall present a one-loop model for long-range $0\nu\beta\beta$ decay. This model only contains two new scalar fields $S_1$, $S_2$ and a new vector-like fermion $F$ which transform under the SM gauge group as
\begin{equation}
\label{eq:new-fields-example-model}S_1\sim\left(1, 1, 1\right),~~~S_2\sim\left( 3, 1, -4/3\right),~~~F\sim\left(1, 1, 2\right)
\end{equation}
in the notation of $\left(SU(3)_C, SU(2)_L, U(1)_Y\right)$. Notice that we assume there is only one generation of the fermion field $F$. Then we can read out the following SM gauge invariant Lagrangian $\mathcal{L}_{int}$ among the SM fields and new fields:
\begin{align}
\nonumber\mathcal{L}_{int}=&y_{I\alpha\beta}\overline{\ell}_{\alpha}i\tau^{2}\ell_{\beta}^{c}S_{1}^{*}
+y_{II\alpha\beta}\overline{e_{R}^{c}}_{,\alpha}d_{R,\beta}S_{2}^{*}+y_{III\alpha}\overline{u_{R}}_{,\alpha}FS_{2}
+y_{IV\alpha}\overline{F}e_{R,\alpha}^{c}S_{1}+\text{h.c.}\\
&+m_{F}\overline{F}F+\sum_{i=1}^{2}m_{S_{i}}^2S_{i}^{\dagger}S_{i}+\sum_{i=1}^{2}\xi_{i}H^\dagger H S_{i}^{\dagger}S_{i}+\sum_{i,j=1}^{2}\zeta_{ij}S_{i}^{\dagger}S_{i} S_{j}^{\dagger}S_{j}\,,
\label{eq:interaction-example-model}
\end{align}
which can generate the one-loop diagram for long-range $0\nu\beta\beta$ decay shown in figure~\ref{fig:example}. We see that the coupling $y_I$ is antisymmetric on the flavor indices, i.e., $y_{I\alpha\beta}=-y_{I\beta\alpha}$. It is important to note that the lepton fields inside the loop can be second or the third generation, while the external lepton fields can only be from the first generation.

After electroweak symmetry breaking, the diagram NL3-1-3-2-7-1-4 reduces to the Feynman diagram displayed in the right panel of figure~\ref{fig:example}. With the interaction Lagrangian in Eq.~\eqref{eq:interaction-example-model}, we can straightforwardly calculate this Feynman diagram, and find that the following effective operator is generated
\begin{align}
\label{eq:onbb-operator-model}\frac{G_{F}}{\sqrt{2}}\epsilon_{V+A}^{V+A}j^{\mu}_{V+A}J_{\mu,V+A}\,,
\end{align}
where $j^{\mu}_{V+A}$ and $J_{\mu,V+A}$ are defined in Eq.~\eqref{eq:L4-fermion}, and the coefficient $\epsilon_{V+A}^{V+A}$ is given by
\begin{equation}
\label{eq:epsilon-V-A}
\epsilon_{V+A}^{V+A}=\frac{\sqrt{2}\,{y_{Ie\tau}y_{II\tau d}y_{IIIu}y_{IVe}}m_{\tau}}{64\pi^{2}G_{F}m_{F}^{3}} \mathcal{D}\left(\frac{m_{\tau}^{2}}{m_{F}^{2}},1,\frac{m_{S_{1}}^{2}}{m_{F}^{2}},\frac{m_{S_{2}}^{2}}{m_{F}^{2}}\right)\,,
\end{equation}
The function $\mathcal{D}(x_{1},x_{2},x_{3},x_{4})$ is the loop integral and it is given by
\begin{align}
\nonumber\mathcal{D}(x_{1},x_{2},x_{3},x_{4})&=\frac{1}{i\pi^{2}}\int d^{4}p\frac{1}{\left(p^{2}-x_{1}\right)\left(p^{2}-x_{2}\right)\left(p^{2}-x_{3}\right)\left(p^{2}-x_{4}\right)}\\
\nonumber&=\frac{x_{1}\left(1-\ln x_{1}\right)}{\left(x_{1}-x_{2}\right)\left(x_{1}-x_{3}\right)\left(x_{1}-x_{4}\right)}+\frac{x_{2}\left(1-\ln x_{2}\right)}{\left(x_{2}-x_{1}\right)\left(x_{2}-x_{3}\right)\left(x_{2}-x_{4}\right)}\\
&+\frac{x_{3}\left(1-\ln x_{3}\right)}{\left(x_{3}-x_{1}\right)\left(x_{3}-x_{2}\right)\left(x_{3}-x_{4}\right)}+\frac{x_{4}\left(1-\ln x_{4}\right)}{\left(x_{4}-x_{1}\right)\left(x_{4}-x_{2}\right)\left(x_{4}-x_{3}\right)}\,.
\end{align}

\subsection{\label{sec:massexample}Prediction for neutrino mass }

\begin{figure}[htbp]
\centering
\includegraphics[width=0.95\textwidth]{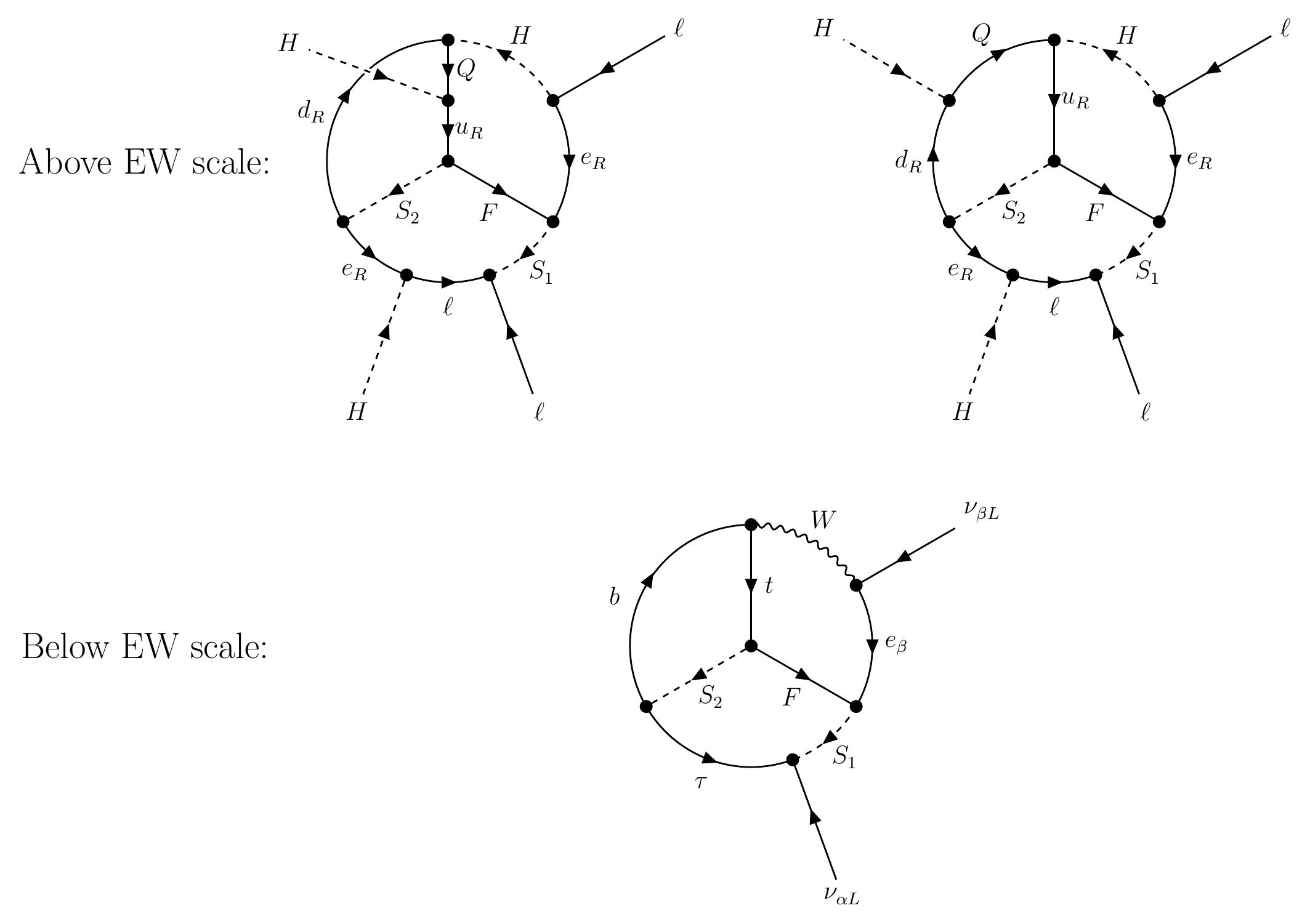}
\caption{The diagrams for the neutrino masses in the example model NL3-1-3-2-7-1-4 of long-range $0\nu\beta\beta$ decay, which can generate non-zero diagonal entries of the neutrino mass matrix. Notice that the two right-handed charged leptons in the top panels can be different flavor.}
\label{fig:MassModelExample}
\end{figure}

\begin{figure}[htbp]
\centering
\includegraphics[width=0.8\textwidth]{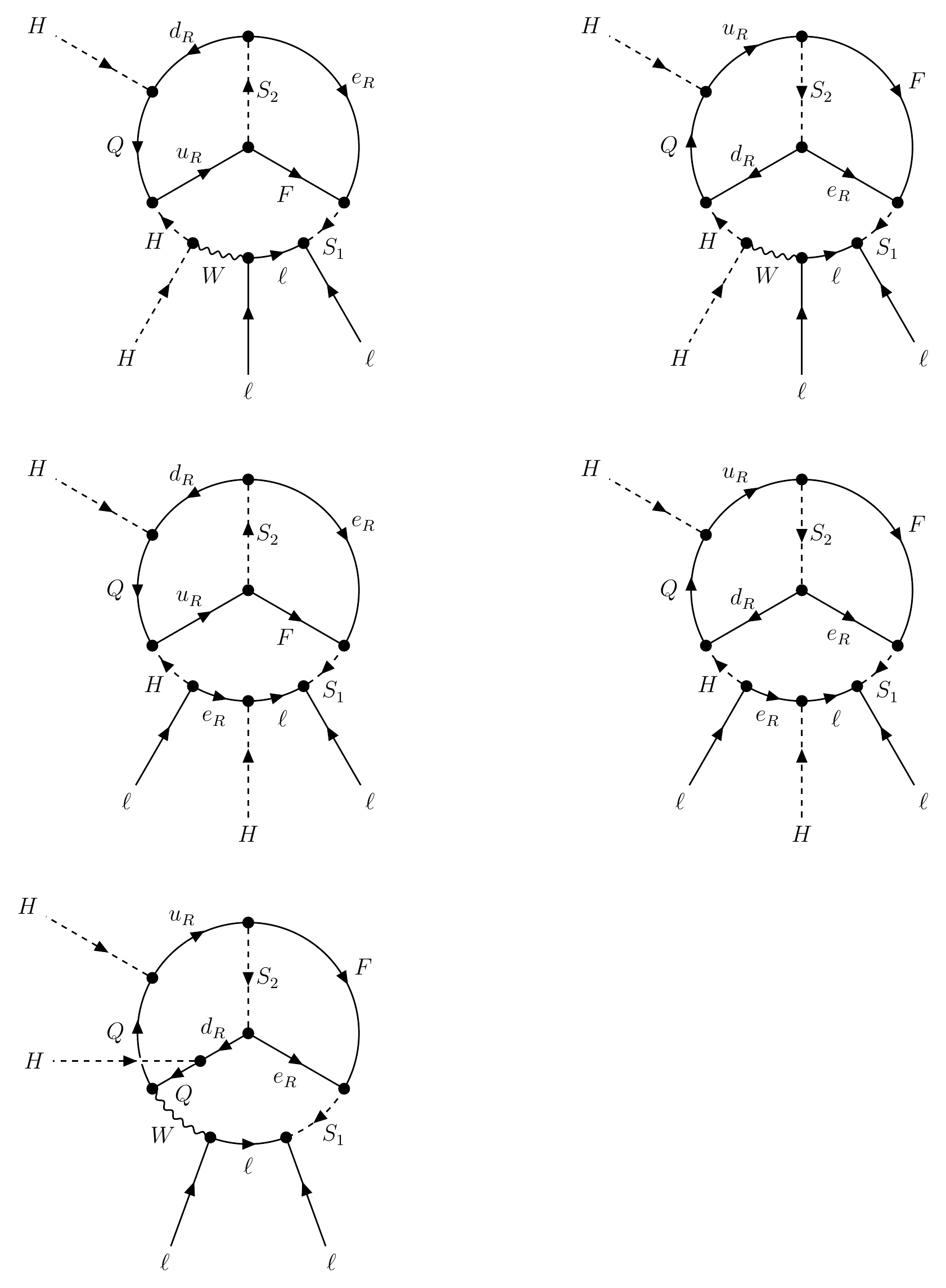}
\caption{The diagrams for the neutrino masses in the example model NL3-1-3-2-7-1-4 of long-range $0\nu\beta\beta$ decay, for which the diagonal entries of the neutrino mass matrix is vanishing. Notice that the inner loops can not be compressed to tree-level renormalizable vertices due to the antisymmetric nature of some $SU(2)_L$ contractions and the chiral structure of the SM.
}
\label{fig:MassModelExample2}
\end{figure}

With the three new fields in Eq.~\eqref{eq:new-fields-example-model} and the relevant interactions, we find that the leading order contributions to the neutrino mass arise at three-loop level, and the corresponding Feynman diagrams are shown in figure~\ref{fig:MassModelExample} and figure~\ref{fig:MassModelExample2}. From the diagram of figure~\ref{fig:MassModelExample} after electroweak symmetry breaking, we see that the neutrino masses are really generated through the black box diagram shown in the right panel of figure~\ref{fig:black_box}. We recall that in the mass mechanism of $0\nu\beta\beta$ decay, the decay amplitude is proportional to the effective Majorana mass $m_{\beta\beta}$ with
\begin{equation}
m_{\beta\beta}=\sum_{i}U_{ei}^{2}m_{i}\,,
\end{equation}
where $U_{ei}$ denote the elements of the neutrino mixing matrix and $m_{i}$ refer to the light neutrino mass. It is known that the effective neutrino mass $m_{\beta\beta}$ is exactly the $(ee)$ entry of the Majorana neutrino mass matrix in the charged lepton diagonal basis.

From figure~\ref{fig:MassModelExample2}, we can see that the two external lepton fields are connected by an internal fermion chain which passes through a flavor off-diagonal interaction vertex associated with $S_1$ (the flavor structure is encoded in the antisymmetric Yukawa $y_I$), while all other vertices on the fermion chain are flavor diagonal. Consequently, the two light neutrinos must have different flavor and the diagonal entries of the neutrino mass matrix is vanishing. However, the diagrams in figure~\ref{fig:MassModelExample} can produce non-zero diagonal elements of the neutrino mass matrix. As a consequence, the contributions of figure~\ref{fig:MassModelExample2} to the $0\nu\beta\beta$ decay via mass mechanism are negligible, and it is sufficient to focus on the diagrams in figure~\ref{fig:MassModelExample}. After the electroweak symmetry breaking, figure~\ref{fig:MassModelExample} produces neutrino masses and the dominant contribution arises from the exchange of the heavy top quark, bottom quark and tau. From the bottom panel of figure~\ref{fig:MassModelExample}, one can straightforwardly calculate the expression of the neutrino mass matrix element as follows
\begin{align}
\nonumber(m_{\nu})_{\alpha\beta}=&\frac{3g^2}{(16\pi^2)^3}V_{tb}\,y^{*}_{II\tau b}y^{*}_{IIIt}\frac{m_{\tau}m_{b}m_{t}m_{e_{\beta}}}{m_{W}^2m_{F}}\\
\label{eq:mnu1-alpha-beta}&\left\{y^{*}_{I\alpha\tau}y^{*}_{IV\beta}\mathcal{M}\left(
\frac{m_{b}^{2}}{m_{F}^2}, \frac{m_{e_\beta}^{2}}{m_{F}^2}, \frac{m_{W}^{2}}{m_{F}^2}, \frac{m_{\tau}^{2}}{m_{F}^2}, \frac{m_{S_{1}}^{2}}{m_{F}^2}, \frac{m_{t}^{2}}{m_{F}^2}, 1, \frac{m_{S_{2}}^{2}}{m_{F}^2} \right)
+(\alpha\leftrightarrow\beta)\right\}\,,
\end{align}
where $g$ is the gauge coupling constant of $SU(2)_L$, $V_{tb}$ is the (33) entry of the Cabibbo-Kobayashi-Maskawa (CKM) quark mixing matrix, and $\mathcal{M}$ denotes a three-loop integral,
\begin{align}
\nonumber&\mathcal{M}\left(x_{1}, x_{2}, x_{3}, x_{4}, x_{5}, x_{6}, x_{7}, x_{8} \right)\\
\nonumber=&\left(\frac{1}{i\pi^2}\right)^{3}\int d^{4}k_{1}d^{4}k_{2}d^{4}k_{3} \left(4x_{3}-k_{2}^{2}\right)\\
        &\frac{1}{k_{1}^{2}-x_{1}} \frac{1}{k_{2}^{2}-x_{2}} \frac{1}{k_{2}^{2}-x_{3}} \frac{1}{k_{3}^{2}-x_{4}} \frac{1}{k_{3}^{2}-x_{5}} \frac{1}{(k_{1}-k_{2})^{2}-x_{6}} \frac{1}{(k_{2}-k_{3})^{2}-x_{7}} \frac{1}{(k_{3}-k_{1})^{2}-x_{8}}\,.
\end{align}
Once the masses of the new fields are specified, the three-loop integrals $\mathcal{M}$ can be computed numerically~\cite{Martin:2016bgz,Freitas:2016zmy}.

\subsection{\label{sec:half-life}Half-life time of $0\nu\beta\beta$ decay }

The neutrinoless double beta decay has been discussed in the framework of effective field theory~\cite{Cirigliano:2017djv}, and the contribution of lepton number violating operator up to dimension seven have been studied. The inverse half-life time of the $0\nu\beta\beta$ decay can be generally expressed as~\cite{Cirigliano:2017djv}
\begin{eqnarray}
\nonumber T_{1/2}^{-1}=&g_{A}^{4}\Bigg\{ G_{01}\abs{\mathcal{A}_{\nu}}^{2}
+4G_{02}\abs{\mathcal{A}_{E}}^{2}+2G_{04}\left[|\mathcal{A}_{m_{e}}|^{2}+\Re\left(\mathcal{A}_{m_{e}}^{*}\mathcal{A}_{\nu}\right)\right] +G_{09}\abs{\mathcal{A}_{M}}^{2}\\
&\qquad-2G_{03}\Re\left(\mathcal{A}_{\nu}\mathcal{A}_{E}^{*}+2\mathcal{A}_{m_{e}}\mathcal{A}_{E}^{*}\right)+G_{06}\Re\left(\mathcal{A}_{\nu}\mathcal{A}_{M}^{*}\right)\Bigg\}\,,
\end{eqnarray}
where $\mathcal{A}_{\nu}$, $\mathcal{A}_{E}$, $\mathcal{A}_{m_{e}}$, $\mathcal{A}_{M}$ depend on nuclear matrix elements and Wilson coefficients of the $\Delta L=2$ operators, and their explicit expressions given in Ref.~\cite{Cirigliano:2017djv} are a bit lengthy. Moreover,  $G_{0i}$ are phase space factors and $g_{A}$ is the well-known unquenched axial coupling, we adopt the values of $G_{0i}$ and $g_{A}$ listed in Ref.~\cite{Cirigliano:2017djv}.

\begin{figure}[!htbp]
\centering
\includegraphics[width=\textwidth]{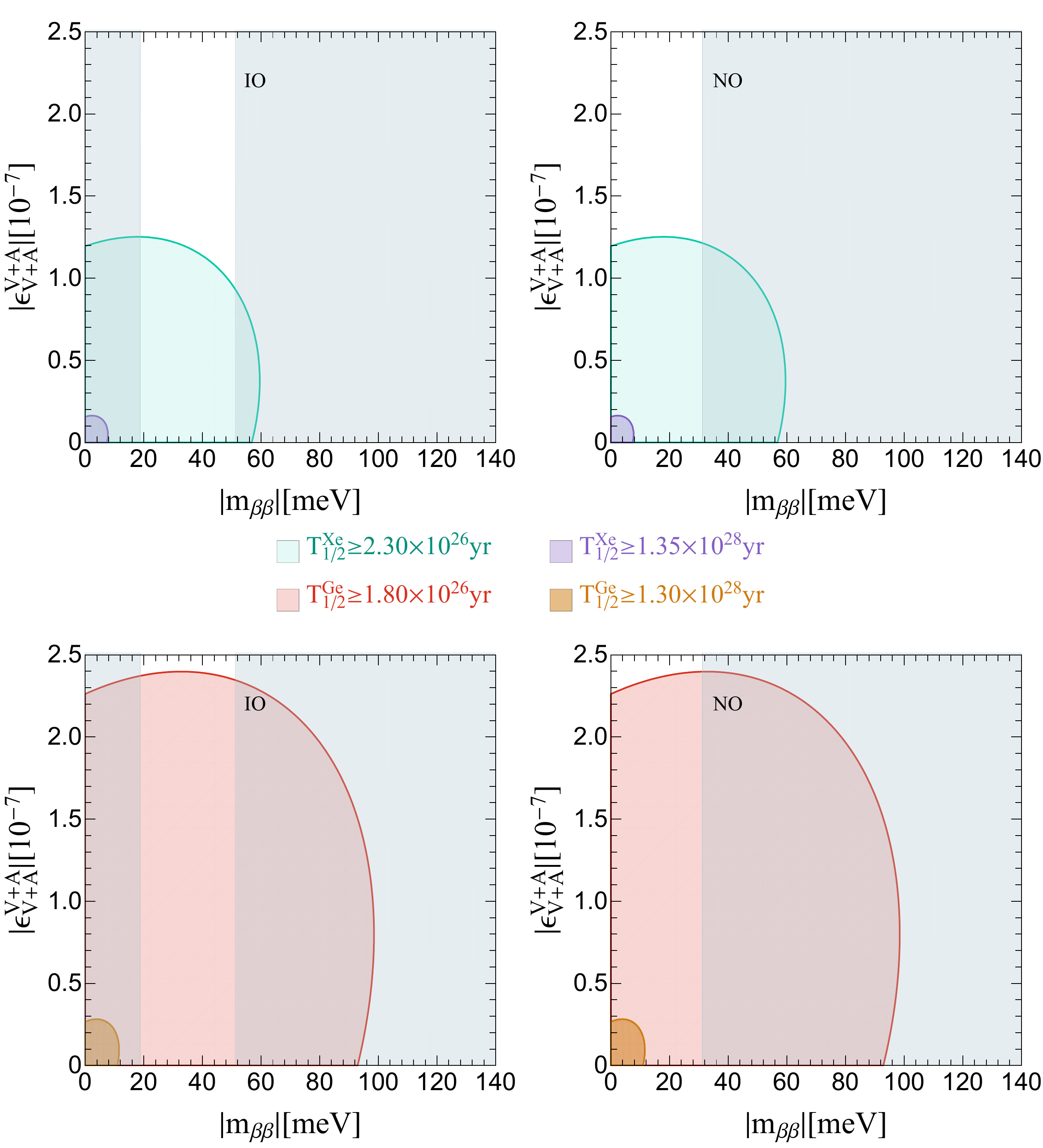}
\caption{\label{fig:halflifePlot} Constraints on the effective neutrino mass $|m_{\beta\beta}|$ and the long-range coupling $|\epsilon_{V+A}^{V+A}|$. In the top panels, the highlighted regions represent the parameter space allowed by the current bounds $T_{1/2}(^{136}\text{Xe})>2.3\times 10^{26}$ yr~\cite{KamLAND-Zen:2022tow} and the future sensitivities $T_{1/2}(^{136}\text{Xe})>1.35\times 10^{28}$ yr~\cite{nEXO:2021ujk}. Similarly the bottom panels display the parameter space allowed by the current bounds $T_{1/2}(^{76}\text{Ge})>1.8\times 10^{26}$ yr~\cite{GERDA:2020xhi} and the future sensitivities $T_{1/2}(^{76}\text{Ge})>1.3\times 10^{28}$ yr~\cite{LEGEND:2021bnm}. The vertical white bands denote the generally allowed region of $m_{\beta\beta}$ for NO and IO neutrino mass spectrum when the experimental data of both neutrino oscillation~\cite{Esteban:2020cvm} and Planck~\cite{Planck:2018vyg} are considered.  }
\end{figure}

As shown in previous section, the mediators of the long-range  $0\nu\beta\beta$ decay model can generate non-vanishing light neutrino masses at three-loop level. Therefore both the long-range mechanism and the mass mechanism contribute to the $0\nu\beta\beta$ decay, and these two contributions should be added coherently. The effective Majorana mass $m_{\beta\beta}$ in mass mechanism leads to non-vanishing $\mathcal{A}_{\nu}$, and the Wilson coefficient $\epsilon_{V+A}^{V+A}$ in Eq.~\eqref{eq:onbb-operator-model} gives rise to the parameters $\mathcal{A}_{E}$ and $\mathcal{A}_{m_{e}}$,
\begin{equation}
\mathcal{A}_{\nu}=\frac{m_{\beta\beta}}{m_{e}}V_{ud}^{2}M_{\nu},~~~\mathcal{A}_{E}=V_{ud}\epsilon_{V+A}^{V+A}M_{E,R},~~~\mathcal{A}_{m_{e}}=V_{ud}\epsilon_{V+A}^{V+A}M_{m_{e},R}\,,
\end{equation}
where $m_{e}$ is the mass of electron, $V_{ud}$ refers to the (11) entry of the CKM matrix, $M_{\nu}$,$M_{E,R}$ and $M_{m_{e},R}$ are the nuclear matrix elements and one should reply on certain nuclear models to calculate their values. Hence the half-life of $0\nu\beta\beta$ decay in our model can be reduced to
\begin{eqnarray}
\nonumber
T_{1/2}^{-1}&=&g_{A}^{4}\Big\{G_{01}\abs{\mathcal{A}_{\nu}}^{2}
+4G_{02}\abs{\mathcal{A}_{E}}^{2}+2G_{04}|\mathcal{A}_{m_{e}}|^{2}\\
&&+2G_{04}|\mathcal{A}_{m_{e}}|\,|\mathcal{A}_{\nu}|\cos\phi
-2G_{03}\left[\abs{\mathcal{A}_{\nu}}\abs{\mathcal{A}_{E}}\cos\phi+2|\mathcal{A}_{m_{e}}|\,|\mathcal{A}_{E}|\right]
\Big\}\,,
\end{eqnarray}
where $\phi$ denotes the relative phase between $m_{\beta\beta}$ and $\epsilon_{V+A}^{V+A}$. We show the constraints of the current and forthcoming $0\nu\beta\beta$ decay experiments on the effective Majorana neutrino mass $|m_{\beta\beta}|$ and the long-range coupling $|\epsilon_{V+A}^{V+A}|$ in figure~\ref{fig:halflifePlot}, where the values of phase space factor and nuclear matrix elements are adopted from Refs.~\cite{Cirigliano:2017djv,Hyvarinen:2015bda}.

It is known that the neutrino mass is tightly constrained by the Planck measurements of the cosmic microwave background anisotropies. Assuming the standard minimal $\Lambda$CDM model and combining with baryon acoustic oscillation measurements, the most stringent bound on neutrino mass is $\sum_{i}m_{\nu_i}<0.12$ eV at 95\% confidence level from the Planck collaboration~\cite{Planck:2018vyg}. Considering the values of the neutrino mass squared differences and mixing angles measured by the neutrino oscillation experiments~\cite{Esteban:2020cvm}, one can obtain that the effective Majorana neutrino mass is in the region: $|m_{\beta\beta}|\leq31.14$ meV for normal ordering (NO) neutrino mass spectrum and $18.72\,\text{meV}\leq|m_{\beta\beta}|\leq51.14$ meV for inverted ordering (IO) neutrino mass, which are shown as vertical white bands in figure~\ref{fig:halflifePlot}.

The highlighted areas in figure~\ref{fig:halflifePlot} denote the allowed regions by the current limits and future sensitivities of the $0\nu\beta\beta$ decay half-life of the isotopes ${}^{76}$Ge and ${}^{136}$Xe, where the relative phase $\phi$ freely varies in the range $0\leq\phi<2\pi$. The next generation tonne-scale experiments of $0\nu\beta\beta$ decay would greatly increase the sensitivity by approximately two orders of magnitude, therefore the constraint on the parameter space would be improved considerably, as can be seen from figure~\ref{fig:halflifePlot}. We notice that the constraint imposed by ${}^{136}$Xe is more stringent than that of ${}^{76}$Ge. It is remarkable that the inverted ordering of neutrino masses may be potentially excluded in the future by next-generation $0\nu\beta\beta$ decay experiments even with the long-range mechanism produced by $d=7$ operators.

\begin{figure}[hptb!]
\centering
\includegraphics[width=0.70\textwidth]{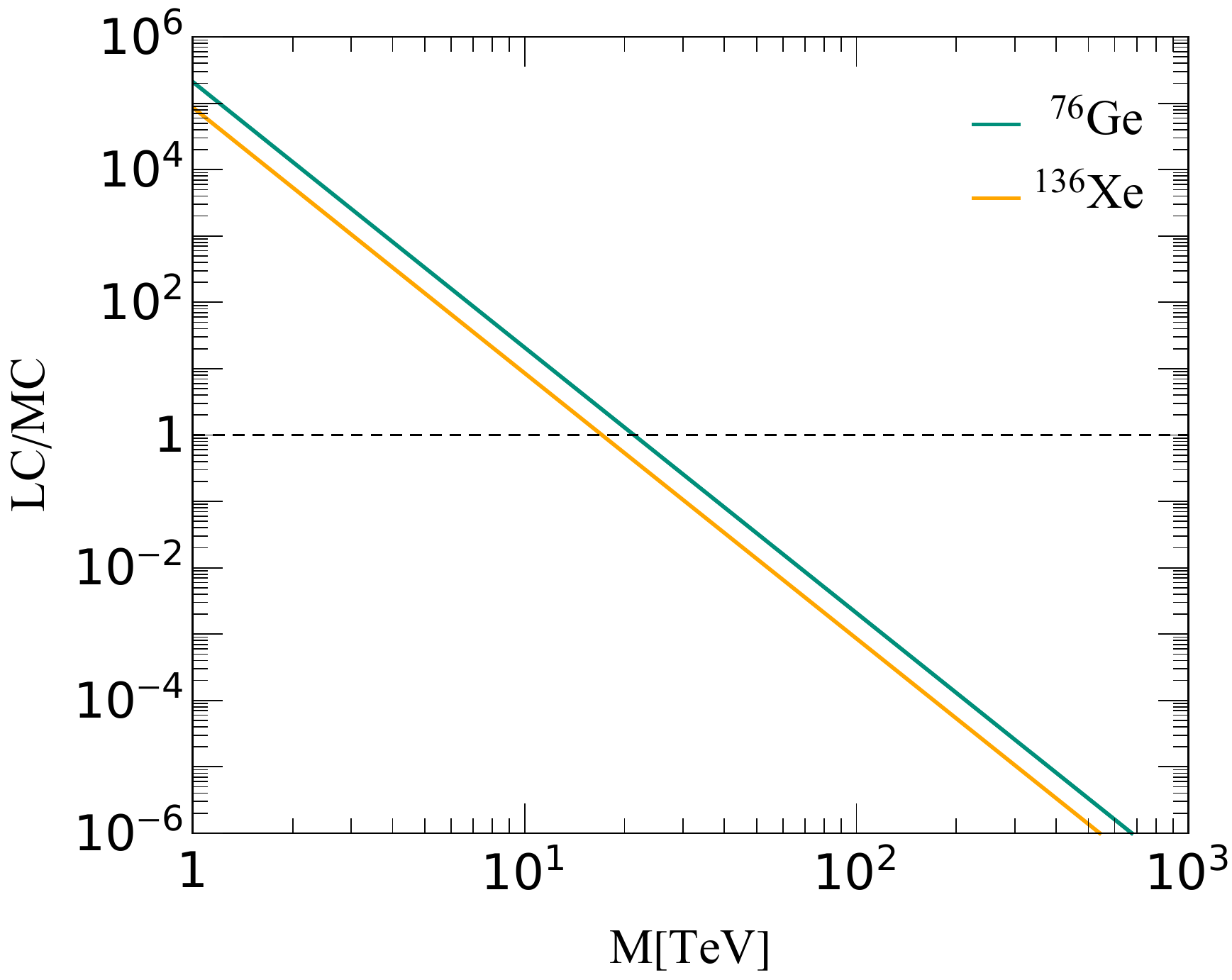}
\caption{The ratio of the long-range mechanism to the mass mechanism. In the figure,
LC$=g_{A}^{4}\left[4G_{02}\abs{\mathcal{A}_{E}}^{2}+2G_{04}|\mathcal{A}_{m_{e}}|^{2}-4G_{03}|\mathcal{A}_{m_{e}}|\,|\mathcal{A}_{E}|\right]$ denotes the long-range contribution to $0\nu\beta\beta$ decay, and MC=$g^4_A G_{01}\abs{\mathcal{A}_{\nu}}^{2}$ is the mass mechanism contribution to $0\nu\beta\beta$ decay. The horizontal dash line, representing $\text{LC/MC}=1$, indicates that the long-range contribution is equal to that of the mass mechanism. The values of phase space factors and nuclear matrix elements used in this figure are taken from~\cite{Cirigliano:2017djv,Hyvarinen:2015bda}.}
\label{fig:dominantPlot}
\end{figure}

We proceed to discuss the relative size of the long-range mechanism and mass mechanism to the decay rate. As an estimation of order of magnitude, we assume that the new fields have the same mass $m_{S_{1}}=m_{S_{2}}=m_{F}=M$ and the new couplings in the Lagrangian of Eq.~\eqref{eq:interaction-example-model} have the same size  $y_{I\alpha\beta}=y_{II\alpha\beta}=y_{III\alpha}=y_{IV\alpha}\equiv y_{eff}$ which are taken to be real. From the expression of the neutrino mass matrix elements given in Eqs.~(\ref{eq:mnu1-alpha-beta}), we see that the effective Majorana mass could scale as
\begin{equation}
m_{\beta\beta}\sim\frac{6g^2}{(16\pi^2)^3}V_{tb}\,y_{eff}^{4}\frac{m_{\tau}m_{b}m_{t}m_{e}}{m_{W}^2M}\,,
\end{equation}
where $M$ is the characteristic mass scale of new particles. Consequently the contribution of the mass mechanism to the $0\nu\beta\beta$ decay is proportional to $y_{eff}^{8}/M^2$, i.e.,
\begin{small}
\begin{align}
        \mathrm{MC}=g^4_A G_{01}\abs{\mathcal{A}_{\nu}}^{2}\sim g^4_AG_{01}\frac{V_{ud}^{4}\abs{M_{\nu}}^{2}}{m_{e}^2}\left[6g^2\frac{V_{tb}y_{eff}^{4}}{(16\pi^2)^3}\frac{m_{\tau}m_{b}m_{t}m_{e}}{m_{W}^2M}\right]^2
        \sim5\times10^{-27}\,\mathrm{yr}^{-1}\cdot y_{eff}^{8}\left(\frac{1\mathrm{GeV}}{M}\right)^{2}\,.
\end{align}
\end{small}
Moreover, the Wilson coefficient $\epsilon_{V+A}^{V+A}$ of the long-range operator in Eq.~\eqref{eq:epsilon-V-A} scales as
\begin{equation}
\epsilon_{V+A}^{V+A}\sim\frac{\sqrt{2}\;y_{eff}^{4}m_{\tau}}{64\pi^{2}G_{F}M^{3}}\,.
\end{equation}
Hence the long-range contribution is proportional to $y_{eff}^{8}/M^6$,
\begin{align}
\nonumber\mathrm{LC}&=g_{A}^{4}\left[4G_{02}|\mathcal{A}_{E}|^{2}+2G_{04}|\mathcal{A}_{m_{e}}|^{2}-4G_{03}|\mathcal{A}_{m_{e}}|\,|\mathcal{A}_{E}|\right]\\
\nonumber&\sim g_{A}^{4}\left(4G_{02}|M_{E,R}|^2+2G_{04}|M_{m_{e},R}|^2-4G_{03}|M_{E,R}|\,|M_{m_{e},R}|\right)V_{ud}^2\left(\frac{\sqrt{2}y_{eff}^{4}m_{\tau}}{64\pi^{2}G_{F}M^{3}}\right)^2\\
&\sim10^{-8}\,\mathrm{yr}^{-1}\cdot y_{eff}^{8}\left(\frac{1\mathrm{GeV}}{M}\right)^{6}\,.
\end{align}
We see that the long-range contribution is comparable to the mass mechanism (i.e., LC/MC$\sim1$) for the new particle mass $M\sim\mathcal{O}(20)$ TeV. The long-range contribution dominates over the mass mechanism with LC/MC$\gg1$ for $M<20$ TeV, while the mass mechanism is dominant in the mass range $M>20$ TeV. We plot the ratio of long-range contribution to mass mechanism with respect to the mass $M$ in figure~\ref{fig:dominantPlot}. We see the ratio LC/MC decreases with the new physics mass scale $M$.

\begin{figure}[hptb!]
\centering
\includegraphics[width=0.99\textwidth]{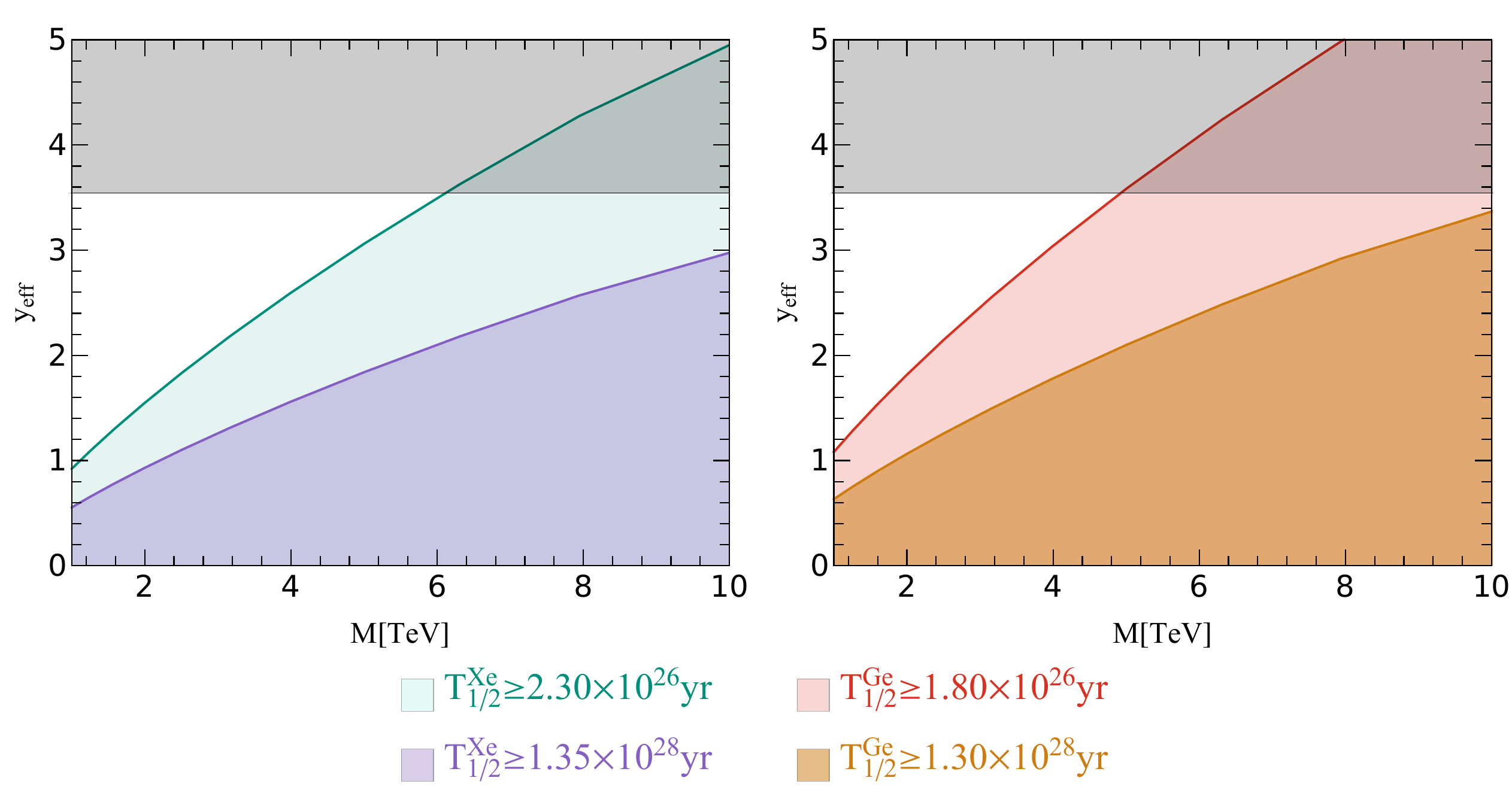}
\caption{The constraint on the effective coupling $y_{eff}$ and new particle mass $M$ by the $0\nu\beta\beta$ decay of ${}^{136}$Xe and ${}^{76}$Ge, where the highlighted regions are allowed by the $0\nu\beta\beta$ decay search. The horizontal grey band is excluded by the perturbative constraint $y_{eff}\leq\sqrt{4\pi}$.}
\label{fig:ybound}
\end{figure}

In figure~\ref{fig:ybound}, we plot the regions of the parameters $y_{eff}$ and $M$ compatible with the current bounds and future sensitivities of $0\nu\beta\beta$ decay search in the isotopes ${}^{136}$Xe and ${}^{76}$Ge. We see that there are still sizable parameter space in which the long-range contribution is dominant.

\section{\label{sec:conclusion}Summary and conclusions }

A lot of neutrino oscillation experiments have established that neutrinos have tiny masses, but the nature of neutrinos is still unknown. If neutrinos are Majorana particles, the light neutrino exchange between two charged current interaction vertices can leads to $0\nu\beta\beta$ decay. This is the so-called mass mechanism, it is not a priori guaranteed to be the dominant contribution in all models. The different possibilities mediating $0\nu\beta\beta$ decay can be generally classified as short-range mechanism, long-rang mechanism and mass mechanism. In the present work, we have performed a systematical decomposition of the dimension-7 long-range $0\nu\beta\beta$ decay operators at one-loop level.

After removing the non-renormalizable topologies and the non-genuine topologies which are one-loop corrections to the tree-level UV completions, we find that there are only 3 genuine one-loop topologies shown in figure~\ref{fig:N0-1}. Subsequently we specify the Lorentz nature of both internal and external fields, these 3 topologies give rise to 8 diagrams in the electroweak basis, as displayed in figure~\ref{fig:DiaN0-1}. In combination with the possible SM quantum number assignments listed in tables~\ref{tab:NL-U1Y},~\ref{tab:NL1-1SU2SU3},~\ref{tab:NL1-2SU2SU3},~\ref{tab:NL1-3SU2SU3} for the each line, one can construct novel one-loop $0\nu\beta\beta$ decay models. Our results can serve as a guide for the construction of one-loop $0\nu\beta\beta$ decay models and for the study of the phenomenology of these models at colliders or high luminosity facilities.

The long-range $0\nu\beta\beta$ decay operators violate lepton number by two unit, consequently the mediators of $0\nu\beta\beta$ decay models generally generate Majorana neutrino masses, as shown in the black box diagram in the right panel of figure~\ref{fig:black_box}. One expects that the long-range $0\nu\beta\beta$ decay models of one-loop can give the dominant contribution to the decay amplitude in certain parameter space if the neutrino mass is generated at two-loop and higher levels, otherwise the long-range contribution should be subdominant to the mass mechanism. Therefore both tree-level and one-loop level contributions to neutrino mass should be forbidden in a genuine one-loop model of long-range $0\nu\beta\beta$ decay, consequently certain SM quantum number assignments for the internal fields are excluded.

Furthermore, we present an example of one-loop $0\nu\beta\beta$ decay model with three-loop neutrino masses. This example requires two new scalar fields and a color singlet vector-like fermion. The predictions for the neutrino mass and $0\nu\beta\beta$ decay are studied. The constraints on the couplings and new physics scale from the $0\nu\beta\beta$ decay search in the isotopes ${}^{76}$Ge and ${}^{136}$Xe are discussed. In this model, the long-range contribution is dominant over the mass mechanism for the new particle mass $M<20$ TeV, while the mass mechanism is dominant in the mass range $M>20$ TeV. Thus we expect this model could be tested at the LHC, or at least LHC can constrain the new messenger fields.

\section*{Acknowledgements}

PTC and GJD are supported by the National Natural Science Foundation of China under Grant Nos.~11975224, 11835013. CYY is supported in part by the Grants No.~NSFC-11975130, No.~NSFC-12035008, No.~NSFC-12047533, the Helmholtz-OCPC International Postdoctoral Exchange Fellowship Program, the National Key Research and Development Program of China under Grant No.~2017YFA0402200, the China Postdoctoral Science Foundation under Grant No.~2018M641621, and the Deutsche Forschungsgemeinschaft (DFG, German Research Foundation) under Germany's Excellence Strategy --- EXC 2121 ``Quantum Universe'' ---390833306.

%\bibliographystyle{utphys}
%\bibliography{references.bib}
%\end{document}

\providecommand{\href}[2]{#2}\begingroup\raggedright\endgroup

\end{document}